\documentclass[aps,twocolumn,preprintnumbers,amsmath,amssymb,nofootinbib,superscriptaddress,notitlepage]{revtex4}
\usepackage{graphicx,color,dcolumn,booktabs,bm}
\usepackage{longtable,lscape}
\usepackage{txfonts}
\usepackage{overpic}
\usepackage{amssymb}
\usepackage{gensymb}
\usepackage{indentfirst}
\usepackage{feynmf}
\usepackage{slashed}
\usepackage{cases}
\usepackage{multirow}
\usepackage{appendix}
\usepackage{float}
\usepackage[figuresright]{rotating}
\usepackage{subfigure}
\usepackage[colorlinks,
citecolor=blue,
anchorcolor=red,
menucolor=red,
linkcolor=red,
filecolor=red,
runcolor=red,
urlcolor=blue,
frenchlinks=red]{hyperref}
\usepackage{epstopdf}
\usepackage{bm}
\usepackage{tabularx}
\usepackage{threeparttable}
\usepackage{bbding}
\usepackage{array}
\usepackage{placeins}
\usepackage{makecell}
\usepackage{comment}
\makeatletter

\newcommand{\Rmnum}[1]{\expandafter\@slowromancap\romannumeral #1@}
\makeatother

\usepackage{soul}

\begin{document}

\title{Understanding the enhanced $\phi \eta^{\prime}$ decay mode of the $\phi(2170)$ through strange-meson loops}

\author{Qin-Song Zhou}
\affiliation{School of Physical Science and Technology, Inner Mongolia University, Hohhot 010021, China}
\affiliation{Research Center for Quantum Physics and Technologies, Inner Mongolia University, Hohhot 010021, China}
\affiliation{Inner Mongolia Key Laboratory of Microscale Physics and Atomic Manufacturing, Hohhot 010021, China}
\affiliation{
Lanzhou Center for Theoretical Physics, Key Laboratory of Theoretical Physics of Gansu Province, Key Laboratory of Quantum Theory and Applications of MoE, Gansu Provincial Research Center for Basic Disciplines of Quantum Physics, Lanzhou University, Lanzhou 730000, China}

\author{Jun-Zhang Wang}\email{wangjzh@cqu.edu.cn}
\affiliation{Department of Physics and Chongqing Key Laboratory for Strongly Coupled Physics, Chongqing University, Chongqing 401331, China}
\author{Dan Guo}\email{guod13@ysu.edu.cn}
\affiliation{Key Laboratory for Microstructural Material Physics of Hebei Province, School of Science, Yanshan University, Qinhuangdao 066004, China}

\date{\today}

\begin{abstract}
The nature of the strangeonium-like $\phi(2170)$ remains controversial. Recent measurements of $e^+e^-\to\phi\eta^{(\prime)}$ reveal a striking puzzle: a broad $\phi(2170)$-like structure appears in the $\phi\eta^\prime$ channel but not in $\phi\eta$, despite the strong phase-space suppression of $\phi\eta^\prime$. We investigate this unexpectedly large $\phi\eta^\prime$ decay fraction within the excited-strangeonium assignment.
A combined analysis of the $e^+e^-\to\phi\eta$ and $e^+e^-\to\phi\eta^\prime$ cross sections is performed by including short-distance amplitudes from vacuum quark-pair creation and long-distance transitions mediated by strange-meson loops. The short-distance mechanism predicts too small a $\phi\eta^\prime/\phi\eta$ strength ratio to explain the data. In contrast, strange-meson loops naturally enhance this ratio because the SU(3)-flavor factor at the $K^{(*)}K^{(*)}\eta^{(\prime)}$ vertex suppresses the transition to $\phi\eta$ relative to $\phi\eta^\prime$. The resulting overall description of both cross sections supports an important role for long-distance dynamics in hidden-strangeness decays of excited vector strangeonia. More intriguingly, although the excited strangeonium contributions have been included,
the current high-precision $\phi\eta$ data still favor the existence of an extra narrow vector state near $2.15~\mathrm{GeV}$ with a width of about $25~\mathrm{MeV}$. If confirmed, it would be a promising exotic-hadron candidate in the light-vector sector. Furthermore, we test $\Gamma(Y\to\phi\eta)/\Gamma(Y\to\phi\eta^\prime)=0.25$, $1$, and $4$ as fit inputs and obtain similarly good descriptions of the cross sections in all three cases. The present data therefore do not allow this ratio to discriminate among different internal configurations of the narrow state.
\end{abstract}

\maketitle

\section{Introduction}\label{section1}

The identification of higher vector $\phi$ mesons remains an open issue in light-hadron spectroscopy. The $\phi(2170)$, also referred to as the $Y(2175)$~\cite{ParticleDataGroup:2026aaa}, has attracted continuous attention since its first observation in the initial-state-radiation process $e^+e^-\to\gamma_{\rm ISR}\phi f_0(980)$~\cite{BaBar:2006gsq}. It has subsequently been studied in many other hidden-strangeness~\cite{Belle:2008kuo,BaBar:2011btv,BESIII:2021aet,BaBar:2007ceh,BESIII:2021bjn,Belle:2022fhh,BESIII:2020gnc} and open-strangeness processes~\cite{BESIII:2018ldc,BESIII:2020vtu,BESIII:2021yam,BESIII:2022wxz,BESIII:2023xac,BaBar:2013jqz,BaBar:2022ahi}. Although the $\phi(1020)$ and $\phi(1680)$ are commonly identified as the $1S$ and $2S$ $s\bar{s}$ vector states, respectively, the nature of the $\phi(2170)$ remains unsettled. Proposed interpretations include conventional excited $s\bar{s}$ states~\cite{Wang:2021gle,Ding:2007pc,Wang:2012wa,Pang:2019ttv,Li:2020xzs,Badalian:2019xir,Feng:2021igh,Hao:2024nvx}, strangeonium hybrids~\cite{Ding:2006ya,Ho:2019org,Ma:2020bex,Guo:2007uz,Li:2025hsp}, fully strange tetraquarks~\cite{Wang:2006ri,Chen:2008ej,Drenska:2008gr,Jiang:2023atq,Deng:2010zzd,Chen:2018kuu,Ke:2018evd,Agaev:2019coa,Liu:2020lpw,Su:2022eun,Xin:2022qnv}, $\phi f_0$ molecular or $\phi K\bar K$ three-body configurations~\cite{MartinezTorres:2008gy,Malabarba:2023zez,Alvarez-Ruso:2009vkn,Oller:2010tr,MartinezTorres:2010ax,Napsuciale:2007wp}, $\Lambda\bar\Lambda$ molecules~\cite{Zhao:2013ffn,Deng:2013aca,Dong:2017rmg}, and triangle-singularity effects~\cite{Wei:2025ejv}. A reliable identification of the $\phi(2170)$ therefore cannot be based on its mass alone, but should also account for correlated information from different decay channels and production mechanisms.

For a long time, the appearance of the $\phi(2170)$ in the $\phi f_0(980)$ mode played a central role in its theoretical interpretation. This prominence of a hidden-strangeness channel motivated exotic pictures such as a $\phi f_0(980)$ hadronic molecule and $\phi K\bar K$ three-body dynamics~\cite{MartinezTorres:2008gy,Malabarba:2023zez}. Recent BESIII measurements of electron-positron annihilation into open-strange channels have provided an important update~\cite{BESIII:2018ldc,BESIII:2020vtu}, with the $\phi(2170)$ also observed in channels such as $K^+K^-$, $K_1(1270)K$, $K_1(1400)K$, and $KK(1460)$. Since open-strange modes are expected to dominate the decays of excited vector $s\bar{s}$ states, these experimental advances make it natural to first examine the $\phi(2170)$ within the excited-strangeonium picture. A useful strategy is to test whether conventional $s\bar{s}$ assignments can coherently describe the available experimental information. If the main features can be accommodated in this framework, the present data do not require an exotic interpretation. If sizable discrepancies remain, the case for nonconventional dynamics becomes more compelling. Along this line, a combined analysis of seven open-strange processes showed that all cross sections can be described simultaneously through interference among the $\phi(3S)$, $\phi(2D)$, and nonresonant amplitudes~\cite{Wang:2021gle}. That analysis also determined the resonance parameters of the relevant $\phi(3S)$ and $\phi(2D)$ states, providing a well-constrained excited-strangeonium baseline for studying additional decay processes associated with the $\phi(2170)$.

Recent measurements of $e^+e^-\to\phi\eta$~\cite{BESIII:2021bjn} and $e^+e^-\to\phi\eta^\prime$~\cite{BESIII:2020gnc} reveal a particularly striking puzzle. BESIII observed a broad structure near $2.18~\mathrm{GeV}$ with a width of about $150~\mathrm{MeV}$ in the $\phi\eta^\prime$ cross section~\cite{BESIII:2020gnc}. This width is consistent with that of the $\phi(2170)$ observed in $\phi f_0(980)$ and multiple open-strange modes~\cite{BESIII:2018ldc,BESIII:2020vtu,BESIII:2021yam,BESIII:2022wxz}. By contrast, the structure reported in $e^+e^-\to\phi\eta$ is much narrower, with a width of only several tens of MeV~\cite{BESIII:2021bjn}, and no evident broad structure appears in the $\phi\eta$ line shape. This difference suggests that the two fitted structures should not be regarded directly as the same resonance contribution. The central puzzle is therefore the unexpectedly large $\phi\eta^\prime$ decay fraction of the broad structure associated with the $\phi(2170)$. The $\phi\eta^\prime$ channel is strongly phase-space suppressed, with phase space favoring $\phi\eta$ by a factor of about $4.3$. This decay pattern might suggest a strangeness-rich configuration inside the $\phi(2170)$, such as a fully strange tetraquark component~\cite{Jiang:2023atq}. A recent theoretical study also emphasized that the ratio of the $\phi\eta$ and $\phi\eta^\prime$ partial widths may constrain the internal structure of the $\phi(2170)$~\cite{Malabarba:2023zez}. That study found that the ratio predicted for a $\phi K\bar K$ molecular state can agree with some experimental solutions within uncertainties. However, as noted above, these experimental solutions should not be interpreted directly as the $\phi\eta^\prime/\phi\eta$ decay ratio of the $\phi(2170)$. This motivates us to analyze the measured $e^+e^-\to\phi\eta$ and $e^+e^-\to\phi\eta^\prime$ cross sections directly, rather than use the resonance parameters and branching fractions extracted separately in the experimental analyses as theoretical inputs.

In this work, we examine whether the unexpectedly large $\phi\eta^\prime$ decay fraction of the $\phi(2170)$ can be understood within the excited-strangeonium picture. We perform a combined analysis of the measured $e^+e^-\to\phi\eta$ and $e^+e^-\to\phi\eta^\prime$ cross sections. Specifically, the $\phi(2170)$ is described as a manifestation of interference between the $\phi(3S)$ and $\phi(2D)$ states, whose resonance parameters are fixed by the previous open-strange analysis~\cite{Wang:2021gle}. The $\phi(1680)$ contribution and nonresonant backgrounds are also included to describe the full line shapes of both processes. The key issue is the mechanism by which these vector strangeonium states decay into the $\phi\eta$ and $\phi\eta^\prime$ final states. We consider short-distance transitions described by quark-pair creation from the vacuum and long-distance transitions induced by strange-meson loops. The former are the direct decay amplitudes of the excited $s\bar{s}$ states, whereas the latter arise from their couplings to open-strange meson pairs followed by rescattering into $\phi\eta^{(\prime)}$.

The comparison of these two mechanisms reveals a clear difference. The short-distance contribution alone predicts a much smaller $\phi\eta^\prime/\phi\eta$ strength and therefore cannot account for the unexpected prominence of the phase-space-suppressed $\phi\eta^\prime$ mode. More importantly, the strange-meson-loop contribution changes this relative strength through the SU(3)-flavor structure of the $K^{(*)}K^{(*)}\eta^{(\prime)}$ vertices, which suppresses the transition to $\phi\eta$ relative to that to $\phi\eta^\prime$. The large $\phi\eta^\prime$ decay fraction can therefore be traced to long-distance strange-meson-loop dynamics within the same excited-strangeonium framework that describes the open-strange line shapes. With both decay mechanisms included, the $e^+e^-\to\phi\eta$ and $e^+e^-\to\phi\eta^\prime$ cross sections can be described well overall. Nevertheless, the $\phi\eta$ data still indicate a possible additional narrow vector structure near $2.15~\mathrm{GeV}$. This structure is a compelling candidate for an exotic state and warrants further high-precision experimental study. We also show that the available cross-section data do not uniquely determine its relative couplings to $\phi\eta^\prime$ and $\phi\eta$. Its $\phi\eta^\prime/\phi\eta$ partial-width ratio alone therefore cannot decisively determine its internal structure.

The remainder of this paper is organized as follows. Section~\ref{SecII} reviews the experimental status of the $\phi(2170)$. Section~\ref{SecIII} presents the theoretical framework for the $e^+e^-\to\phi\eta^{(\prime)}$ cross sections. Section~\ref{SecIV} gives the combined analysis of the $\phi\eta$ and $\phi\eta^\prime$ cross sections and discusses the dynamical origin of the enhanced $\phi\eta^\prime$ decay fraction of the $\phi(2170)$. Section~\ref{SecV} examines the possible additional narrow vector structure and the implications of its $\phi\eta^\prime/\phi\eta$ partial-width ratio. Section~\ref{SecVI} provides the summary.

\section{Experimental status of the $\phi(2170)$}
\label{SecII}

The present understanding of the $\phi(2170)$ relies on a broad set of
measurements rather than on a single discovery channel.  It is therefore
necessary to review the relevant experimental information in more detail before
turning to the analysis of the $\phi\eta^{(\prime)}$ cross sections.  

\begin{figure*}[!t]
  \centering
  \includegraphics[width=\textwidth]{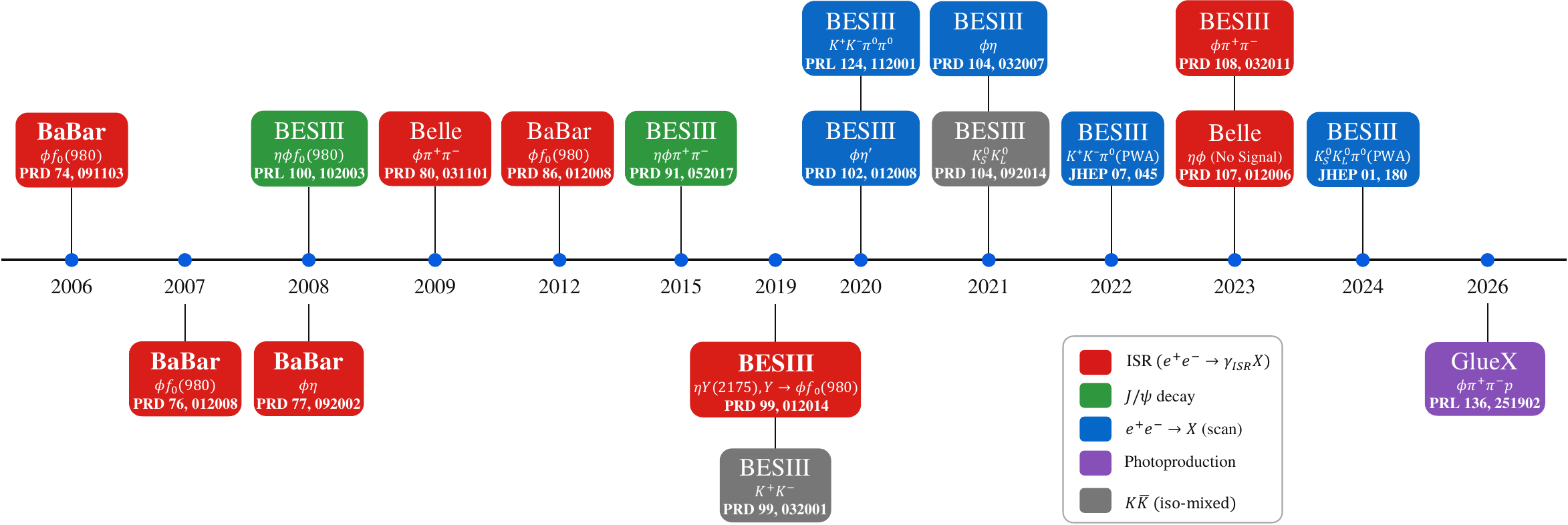}
  \caption{{Timeline of experimental studies of the $\phi(2170)$.}}\label{Fig:expstatus}
\end{figure*}

The vector state denoted the $Y(2175)$ is currently listed by the Particle Data Group (PDG) as the $\phi(2170)$, with quantum numbers $I^{G}(J^{PC})=0^{-}(1^{--})$~\cite{ParticleDataGroup:2026aaa}. 
The 2026 PDG averages of its resonance parameters are $m=(2164\pm5)~\mathrm{MeV}$ and $\Gamma=88^{+26}_{-21}~\mathrm{MeV}$. The uncertainty on the width is large because the individual measurements are mutually inconsistent~\cite{ParticleDataGroup:2026aaa}. This already points to substantial channel dependence among the reported resonance parameters associated with $\phi(2170)$.

The first evidence was reported by BaBar in 2006 using the initial-state-radiation (ISR) process $e^{+}e^{-}\to\gamma_{\rm ISR}\phi f_{0}(980)$ with $\phi\to K^{+}K^{-}$ and $f_{0}(980)\to\pi\pi$~\cite{BaBar:2006gsq}. The fitted mass and width were $m=(2175\pm10\pm15)~\mathrm{MeV}, \Gamma=(58\pm16\pm20)~\mathrm{MeV}$. Subsequent BaBar analyses of the charged- and neutral-dipion modes gave compatible masses near $2.17$--$2.19~\mathrm{GeV}$, although these results were based on closely related data samples~\cite{BaBar:2007ptr}.
BaBar also investigated the $\phi\eta$ channel and obtained a lower mass, $2125\pm22\pm10~\mathrm{MeV}$, albeit with limited statistical precision~\cite{BaBar:2007ceh}. Independently, BESII observed an enhancement in $J/\psi\to\eta Y$, $Y\to\phi f_{0}(980)$ with $m=(2186\pm10\pm6)~\mathrm{MeV}$ and $\Gamma=(65\pm23\pm17)~\mathrm{MeV}$, providing the first confirmation through a production mechanism other than direct $e^{+}e^{-}$ annihilation~\cite{BES:2007sqy}.

In 2009, Belle studied the ISR processes $e^{+}e^{-}\to\gamma_{\rm ISR}\phi\pi^{+}\pi^{-}$ and $\gamma_{\rm ISR}\phi f_{0}(980)$~\cite{Belle:2008kuo}. {The resulting resonance parameters, $m=2079\pm13^{+79}_{-28}~\mathrm{MeV}$ and $\Gamma=192\pm23^{+25}_{-61}~\mathrm{MeV}$, corresponded to a lower mass and considerably broader structure than those observed by BaBar and BESII.} The large asymmetric model uncertainties, however, substantially limit the interpretation of this result. A higher-statistics BaBar analysis subsequently obtained $m=(2180\pm8\pm8)~\mathrm{MeV}, \Gamma=(77\pm15\pm10)~\mathrm{MeV}$ from the $\phi f_{0}(980)$ line shape, including interference with the $\phi(1680)$~\cite{BaBar:2011btv}. BESIII later observed the state with a significance exceeding $10\sigma$ in $J/\psi\to\eta\phi\pi^{+}\pi^{-}$, obtaining $m=(2200\pm6\pm5)~\mathrm{MeV}, \Gamma=(104\pm15\pm15)~\mathrm{MeV}$~\cite{BESIII:2014ybv}. 

A different production mechanism was investigated by BESIII through $e^{+}e^{-}\to\eta Y(2175), Y(2175)\to\phi f_{0}(980)$ at center-of-mass energies between $3.7$ and $4.6~\mathrm{GeV}$~\cite{BESIII:2017qkh}. {The measured resonance parameters, $m=(2135\pm8\pm9)~\mathrm{MeV}$ and $\Gamma=(104\pm24\pm12)~\mathrm{MeV}$, gave a mass lower than those obtained in most direct-production measurements in electron-positron collisions.} In the same period, a structure was observed in the $e^{+}e^{-}\to K^{+}K^{-}$ cross section with $m=(2239.2\pm7.1\pm11.3)~\mathrm{MeV}, \Gamma=(139.8\pm12.3\pm20.6)~\mathrm{MeV}$~\cite{BESIII:2018ldc}. This enhancement is also referred to as the $Y(2240)$. Since the electromagnetic $K\bar K$ amplitude can contain both isoscalar and isovector components, it may include contributions from the $\phi(2170)$, $\omega$-like and $\rho$-like states, and is therefore not included in the PDG averages of  $\phi(2170)$. 
{For the $K\bar K$ channel, Ref.~\cite{Wang:2021gle} calculated the contributions from the nearby excited $\rho$, $\omega$, and $\phi$ states by combining their mass spectra, $K\bar K$ decay widths, and dileptonic couplings. The structure around $2.2~\mathrm{GeV}$ was found to be dominated by the interference structure from the $\phi(3S)$ and $\phi(2D)$ contributions.}

The BESIII energy scan data have enabled measurements in several additional hidden- and open-strangeness channels. A partial-wave analysis of $e^{+}e^{-}\to K^{+}K^{-}\pi^{0}\pi^{0}$ gave $m=(2126.5\pm16.8\pm12.4)~\mathrm{MeV}, \Gamma=(106.9\pm32.1\pm28.1)~\mathrm{MeV}$~\cite{BESIII:2020vtu}. 
{In the $\phi\eta'$ channel, BESIII observed a prominent structure with $m=(2177.5\pm4.8\pm19.5)~\mathrm{MeV}$ and $\Gamma=(149.0\pm15.6\pm8.9)~\mathrm{MeV}$~\cite{BESIII:2020gnc}. Its mass is close to the nominal $\phi(2170)$ value, but its width is substantially larger than those obtained in the early measurements of the $\phi f_{0}(980)$ channel.} By contrast, the $e^{+}e^{-}\to\phi\eta$ measurement yielded one of the narrowest reported line shapes, $m=(2163.5\pm6.2\pm3.0)~\mathrm{MeV}, \Gamma=\left(31.1^{+21.1}_{-11.6}\pm1.1\right)~\mathrm{MeV}$~\cite{BESIII:2021bjn}.

A structure near $2.27~\mathrm{GeV}$ was also observed in $e^{+}e^{-}\to K_{S}^{0}K_{L}^{0}$, with $m=(2273.7\pm5.7\pm19.3)~\mathrm{MeV}, \Gamma=(86\pm44\pm51)~\mathrm{MeV}$~\cite{BESIII:2021yam}. {Its mass is compatible within uncertainties with that of the $Y(2240)$ observed in the $K^{+}K^{-}$ channel, whereas the width has a smaller central value, although the two widths remain consistent within the large uncertainties.} A partial-wave analysis of $e^{+}e^{-}\to K^{+}K^{-}\pi^{0}$, dominated by the $K^{*}(892)^{+}K^{-}$ and $K_{2}^{*}(1430)^{+}K^{-}$ intermediate states, gave $m=(2190\pm19\pm37)~\mathrm{MeV}, \Gamma=(191\pm28\pm60)~\mathrm{MeV}$~\cite{BESIII:2022wxz}. This is one of the broadest measurements included in the current PDG average. {The high-statistics BESIII energy scan of $e^{+}e^{-}\to\phi\pi^{+}\pi^{-}$ obtained $m=(2178\pm20\pm5)~\mathrm{MeV}$ and $\Gamma=(140\pm36\pm16)~\mathrm{MeV}$ in the $\phi f_{0}(980)$-enriched sample~\cite{BESIII:2021aet}. This result favors a broad $\phi f_{0}(980)$ line shape, consistent with the broad structure associated with the $\phi(2170)$ in $\phi\eta^\prime$ and multiple open-strange modes. The smaller widths obtained in the early ISR measurements of the $\phi f_{0}(980)$ channel are likely due to their limited precision. The BESIII energy scan instead favors a broad line shape.} {In the $\phi\eta$ channel, Belle found no significant $\phi(2170)$ contribution in its higher-statistics ISR measurement~\cite{Belle:2022fhh}. When the mass and width of the narrow structure reported by BESIII were fixed in the fit, its significance was only $1.7\sigma$.}

The most recent BESIII result included in the 2026 PDG average is the partial-wave analysis of $e^{+}e^{-}\to K_{S}^{0}K_{L}^{0}\pi^{0}$. Evidence for a structure in the $K^{*}(892)^{0}\bar K^{0}$ line shape was found with $m=(2164.7\pm9.1\pm3.1)~\mathrm{MeV}, \Gamma=(32.4\pm21.0\pm1.8)~\mathrm{MeV}$
\cite{BESIII:2023xac}. Its mass and width are consistent with the narrow structure observed in the $\phi\eta$ channel, although the statistical uncertainty remains large. 

Complementary information has recently been obtained from photoproduction. GlueX searched for the $\phi(2170)$ state in
$\gamma p\to\phi\pi^{+}\pi^{-}p$.
 A line shape fixed to the nominal $\phi(2170)$ parameters showed a significance of only $1.9\sigma$ after systematic uncertainties were included, together with the upper limit $\sigma[\gamma p\to\phi(2170)p]\,\mathcal{B}[\phi(2170)\to\phi\pi^{+}\pi^{-}]<272~\mathrm{pb}$ at the $90\%$ confidence level~\cite{GlueX:2025dgj}.
A free fit gave
$m=(2249\pm12\pm10)~\mathrm{MeV}, \Gamma=(153\pm54\pm76)~\mathrm{MeV}$, suggesting that the higher-mass enhancement structure may be distinct from the nominal $\phi(2170)$, although a dedicated partial-wave analysis is still required.

Several general features emerge from the existing measurements. First, the fitted mass is more stable than the width in channels containing a $\phi(2170)$-like state. The peak positions of most such measurements cluster between $2.16$ and $2.20~\mathrm{GeV}$, whereas the principal lower-mass results occur near $2.12$--$2.14~\mathrm{GeV}$. The structures observed in the $K\bar K$ channels and in GlueX photoproduction instead tend to lie near $2.24$--$2.27~\mathrm{GeV}$.

Second, the reported Breit--Wigner widths exhibit a strong channel dependence. {The $\phi\eta$ and neutral $K^{*}(892)^{0}\bar K^{0}$ measurements give narrow widths of about $30~\mathrm{MeV}$. The $\phi\eta^\prime$, $\phi\pi^{+}\pi^{-}$, and charged $K^{*}(892)^{+}K^{-}$ analyses instead favor widths of $140$--$190~\mathrm{MeV}$, while the early $\phi f_{0}(980)$ measurements lie between these two groups.} The six measurements entering the PDG width average have $\chi^{2}=30.1$, corresponding to a confidence level below $10^{-4}$ \cite{ParticleDataGroup:2026aaa}. {The PDG value therefore summarizes a set of channel-dependent Breit--Wigner widths and should not be interpreted as a universal resonance width. Such a pattern can arise when several nearby vector states contribute
to different channels with channel-dependent relative strengths and
interference phases, so that each cross section exhibits only a single
visible structure but with a distinct line shape. This provides a
natural explanation for the channel dependence of the fitted
$\phi(2170)$ parameters. The analysis of open-strange channels in
Ref.~\cite{Wang:2021gle}, which includes interference between the
$\phi(3S)$ and $\phi(2D)$ states, provides a concrete example.} 

{In summary, the $\phi(2170)$ has been investigated in ISR production, direct $e^{+}e^{-}$ energy scans, $J/\psi$ decays, and photoproduction, as summarized in Fig.~\ref{Fig:expstatus}. Most observed structures associated with $\phi(2170)$ in various final states  have masses between $2.16$ and $2.20~\mathrm{GeV}$, while lower-mass structures near $2.12$--$2.14~\mathrm{GeV}$ and higher-mass enhancements near $2.24$--$2.27~\mathrm{GeV}$ have also been reported. The fitted widths span approximately $30$--$190~\mathrm{MeV}$. Recent BESIII measurements have substantially extended the experimental information in both hidden- and open-strange channels, but sizeable differences among the reported resonance parameters remain.}

\section{Theoretical framework}\label{SecIII}

\subsection{Cross section formula for $e^+e^-\to\phi\eta^{(\prime)}$}
{In general, the $e^+e^-\to\phi\eta^{(\prime)}$ cross section receives two types of contributions: a direct continuum background and resonant amplitudes from intermediate strangeonium-like states. It can therefore be modeled as their coherent sum~\cite{ParticleDataGroup:2026aaa}:}
\begin{eqnarray}
\label{Eq1} 
\sigma_f(s)=\left|\mathcal{M}^{\rm Con}_f(s)+\sum_k e^{i\varphi_{fk}}\mathcal{M}_{k}(s)\right|^2.
\end{eqnarray}
{Here, $\varphi_{fk}$ is the relative phase between the $k$th resonant amplitude and the continuum amplitude, and $f=1$ ($2$) denotes the $\phi\eta$ ($\phi\eta^\prime$) channel. The continuum amplitude $\mathcal{M}_f^{\rm{Con}}$ is parametrized as}
\begin{eqnarray}\label{Eq2}
\mathcal{M}^{\rm{Con}}_f=a_f \cdot e^{-b_f(E_{\rm{cm}}-m_{\phi}-m_{\eta^{(\prime)}})}\sqrt{\Phi_2(s)},
\end{eqnarray}
{where $\Phi_2(s)$ is the two-body phase-space factor, and $a_f$ and $b_f$ are free parameters that govern the normalization and energy dependence of the continuum amplitude, respectively. The contribution from an intermediate strangeonium-like state $R_k$ is described by a relativistic Breit--Wigner amplitude~\cite{ParticleDataGroup:2024cfk}:}
\begin{eqnarray}\label{Eq3}
\mathcal{M}_{k}=\frac{\sqrt{12\pi\Gamma^{e^+e^-}_{R_k}\mathcal{B}(\phi\eta^{(\prime)})\Gamma^{\rm{tot}}_{R_k}}}{s-M_{R_k}^2+iM_{R_k}\Gamma_{R_k}^{\rm{tot}}}\sqrt{\frac{\Phi_2(s)}{\Phi_2(M_{R_k}^2)}}.
\end{eqnarray}
{$M_{R_k}$, $\Gamma_{R_k}^{\rm{tot}}$, and $\Gamma_{R_k}^{e^+e^-}$ denote the mass, total width, and dileptonic width of the $k$th intermediate state, respectively, while $\mathcal{B}(\phi\eta^{(\prime)})$ is its branching fraction to $\phi\eta^{(\prime)}$. Throughout this work, $\eta^{(\prime)}$ denotes either $\eta$ or $\eta^{\prime}$.}

\subsection{The mechanisms of excited strangeonium decays to $\phi\eta^{(\prime)}$}

\begin{figure}[!htb]
  \centering
  \includegraphics[width=250pt]{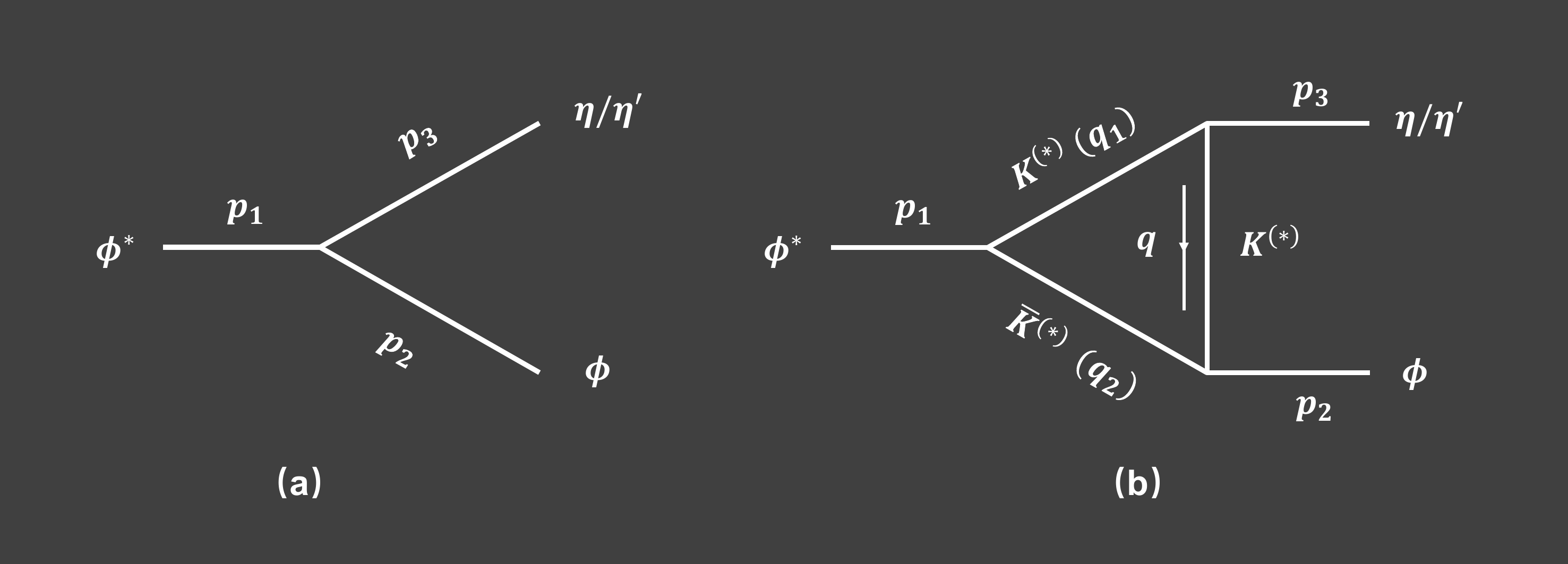}
  \caption{Two decay mechanisms for highly excited strangeonium states into $\phi\eta^{(\prime)}$. (a) OZI-allowed tree-level process. (b) Loop-induced transition through intermediate $K^{(*)}\bar K^{(*)}$ loops.}\label{F1}
\end{figure}

In this work, we consider two decay mechanisms for the excited strangeonium states into $\phi\eta^{(\prime)}$, as illustrated in Fig.~\ref{F1}. Figure~\ref{F1}~(a) represents the OZI-allowed tree-level contribution, where an excited strangeonium state directly decays into $\phi\eta^{(\prime)}$ via the quark-pair-creation (QPC) mechanism. Figure~\ref{F1}~(b) depicts the long-distance rescattering contribution, in which the initial strangeonium state couples to intermediate $K^{(*)}\bar{K}^{(*)}$ meson pairs, followed by their rescattering into the hidden-strange final state $\phi\eta^{(\prime)}$.
{Hadronic loop mechanism had been widely employed in studies of hidden-flavor transitions of excited charmonia and bottomonia~\cite{Bai:2026atm}. A representative example is the hadronic-loop prediction that the branching fractions of $\Upsilon(5S)\to\eta\Upsilon_J(1D)$ can be of order $10^{-3}$~\cite{Wang:2016qmz}, which was subsequently confirmed by Belle~\cite{Belle:2018hjt}.}

Here, we employ an effective Lagrangian approach to describe the interaction vertices involved in Fig.~\ref{F1}. The relevant effective Lagrangians are given by~\cite{Kaymakcalan:1983qq,Zhou:2022wwk,Zhou:2022ark}
\begin{eqnarray}
\label{Eq4}
\mathcal{L}_{\phi^{(*)}\phi\eta^{(\prime)}}&=&g_{\phi^*\phi\eta^{(\prime)}}\varepsilon_{\mu\nu\alpha\beta}  \partial^{\mu}\phi^{*\nu}\partial^{\alpha}\phi^{\beta}\eta^{(\prime)}\\ 
\nonumber\\
\mathcal{L}_{\phi K^{(*)}K^{(*)}}&=&-ig_{\phi K K}\phi^{\mu}(\bar{K}\overleftrightarrow{\partial}_{\mu}K)\nonumber\\
&&-g_{\phi K^* K}\epsilon^{\mu\nu\alpha\beta}\partial_\mu\phi_{\nu} (\bar{K} \partial_\alpha K_{\beta}^*+\partial_\alpha \bar{K}_{\beta}^* K) \nonumber\\
&&-ig_{\phi K^* K^*}(\bar{K}_{\nu}^* \overleftrightarrow{\partial}_{\mu} K^{*\nu} \phi^\mu+ \bar{K}^{*\mu} K_{\nu}^* \overleftrightarrow{\partial}_{\mu} \phi^{\nu}\nonumber\\ \label{Eq5}
 &&+ \phi_{\nu} \overleftrightarrow{\partial}_{\mu} \bar{K}^{*\nu} K^{*\mu} ),\\ 
 \nonumber\\
\mathcal{L}_{K^{(*)}K^{(*)}\eta^{(\prime)}}&=& ig_{K^*K\eta^{(\prime)}}\left[(\bar{K}^{*\mu}K\partial_{\mu}\eta^{(\prime)}-\bar{K}^{*\mu}\partial_{\mu}K\eta^{(\prime)})\right.\nonumber\\
&&\left.+(\partial_{\mu}\bar{K}K^{*\mu}\eta^{(\prime)}-\bar{K}K^{*\mu}\partial_{\mu}\eta^{(\prime)})\right]\nonumber\\ \label{Eq6}
&&+g_{K^*K^*\eta^{(\prime)}}\varepsilon_{\mu\nu\alpha\beta}  \partial^{\mu}\bar{K}^{*\nu}\partial^{\alpha}K^{*\beta}\eta^{(\prime)},
\end{eqnarray}
{Here, $\phi^{*}$, $\phi$, $\eta^{(\prime)}$, and $K^{(*)}$ denote an excited vector strangeonium, the ground state $\phi(1020)$, the $\eta^{(\prime)}$, and the kaon fields, respectively.}

{Using these effective Lagrangians, the tree-level amplitude for the process in Fig.~\ref{F1}~(a) is}
\begin{eqnarray}
\mathcal{M}_{\rm{Tree}}&=&g_{\phi^*_i\phi\eta^{(\prime)}}\varepsilon^{\mu}(p_1)\epsilon_{\nu\mu\alpha\beta}(-i p_1^{\nu})(ip_2^{\alpha})\varepsilon^{*\beta}(p_2). \label{Eq7}
\end{eqnarray}
{The loop amplitude for the process in Fig.~\ref{F1}~(b) receives contributions from the six allowed intermediate $K^{(*)}\bar{K}^{(*)}$ configurations shown in Appendix~\ref{appendix}:}
\begin{eqnarray}
\mathcal{M}_{\rm{Loop}}&=&4\sum_{i=1}^{6}\mathcal{M}_{\rm{Loop}}^{(i)}, \label{Eq8}
\end{eqnarray}
{The factor of $4$ accounts for the charge-conjugate and isospin-related intermediate kaon loops. Explicit expressions for $\mathcal{M}_{\mathrm{Loop}}^{(i)}$ are given in Eqs.~\eqref{eqA1}--\eqref{eqA6} of the Appendix. To account for the off-shell effect of the exchanged meson and regulate the ultraviolet behavior of the loop integral, we introduce the phenomenological dipole form factor $\mathcal{F}(q,m_{K^{(*)}}^{2})$~\cite{Locher:1993cc,Li:1996yn,Cheng:2004ru}:}
\begin{eqnarray}
\mathcal{F}(q,m_{K^{(*)}}^{2})=\left(\frac{m_{K^{(*)}}^{2}-\Lambda^2}{q^{2}-\Lambda^2}\right)^2. \label{Eq9}
\end{eqnarray}
{The cutoff is parametrized as $\Lambda=m_{K^{(*)}}+\alpha_{\Lambda}\Lambda_{\rm{QCD}}$, with $\Lambda_{\rm{QCD}}=0.22~\mathrm{GeV}$. The dimensionless parameter $\alpha_\Lambda$, which is usually taken to be in the range 1--5~\cite{Cheng:2004ru,Wang:2022jxj,Peng:2024xui,Qian:2023taw}, is the only free parameter in the decay amplitudes. The total amplitude for $\phi^{*}\to\phi\eta^{(\prime)}$ is then}
\begin{eqnarray}
\mathcal{M}_{\rm{Total}}=\mathcal{M}_{\rm{Tree}}+{\mathcal{M}_{\rm{Loop}}}. \label{Eq10}
\end{eqnarray}

{The branching fraction for an excited strangeonium state decaying into $\phi\eta^{(\prime)}$ is}
\begin{eqnarray}
\mathcal{B}(\phi^*\to\phi\eta^{(\prime)})=\frac{1}{3}\frac{1}{\Gamma^{\rm{tot}}}\frac{\lvert \vec{p}_2 \rvert}{8\pi m_1^2}\lvert\mathcal{M}_{\rm{Total}}\rvert^2, \label{Eq11}
\end{eqnarray}
{Here, the factor $1/3$ averages over the polarization of the initial state, and $\vec{p}_2$ is the momentum of the final-state $\phi$ meson in the rest frame of the decaying $\phi^{*}$.}

{To evaluate these branching fractions, we first determine the coupling constants in the decay amplitudes. The value $g_{\phi KK}=4.57$ is extracted from the measured partial width of $\phi(1020)\to K^{+}K^{-}$~\cite{ParticleDataGroup:2026aaa}. In the SU(3)-flavor limit, the remaining ground-state couplings are related to $g_{\phi KK}$ as follows~\cite{Gilman:1987ax}:}
\begin{eqnarray}\nonumber
g_{\phi K^*K}&=&g_{VVP},  \quad \quad \quad  \quad g_{\phi K^*K^*}=g_{\phi KK},\\ \label{Eq12}
g_{K^*K\eta}&=&(\alpha+\gamma)g_{\phi K K},\quad g_{K^*K\eta^{\prime}}=(\beta+\delta)g_{\phi K K}, \\ \nonumber
g_{K^*K^*\eta}&=&(\alpha+\gamma)g_{VVP},\quad g_{K^*K^*\eta^{\prime}}=(\beta+\delta)g_{VVP}.\\ \nonumber
\end{eqnarray}
{Here, $g_{VVP}=3g_{\phi KK}^2/(4\pi^2f_\pi)$ and the pion decay constant is $f_\pi=132~\mathrm{MeV}$~\cite{Kaymakcalan:1983qq}. The parameters $\alpha$, $\beta$, $\gamma$, and $\delta$, which depend on the $\eta_1$--$\eta_8$ mixing angle, are defined as}
\begin{eqnarray}
\alpha&=&\frac{\cos\theta-\sqrt{2}\sin\theta}{\sqrt{6}},\quad\quad\beta=\frac{\sqrt{2}\cos\theta+\sin\theta}{\sqrt{6}},\nonumber\\ \label{Eq13}
\\ \nonumber
\gamma&=&\frac{-2\cos\theta-\sqrt{2}\sin\theta}{\sqrt{6}},\quad\delta=\frac{\sqrt{2}\cos\theta-2\sin\theta}{\sqrt{6}}.
\end{eqnarray}

For the mixing angle $\theta=-14.1^\circ$~\cite{Christ:2010dd,Yu:2025pyu}, these relations yield
\begin{eqnarray}
\label{Eq14}
\frac{g_{K^{(*)}K^{(*)}\eta}}
     {g_{K^{(*)}K^{(*)}\eta^\prime}}=\frac{\alpha+\gamma}{\beta+\delta}=\frac{-\cos\theta-2\sqrt{2}\sin\theta}
       {2\sqrt{2}\cos\theta-\sin\theta}\simeq -0.094 .
\end{eqnarray}
{This small coupling ratio suppresses the $\phi\eta$ loop amplitude by about one order of magnitude relative to the $\phi\eta^\prime$ amplitude, corresponding to roughly two orders of magnitude at the rate level. The triangle  loop mechanism can therefore overcome the smaller phase space of $\phi\eta^\prime$, in marked contrast to the QPC amplitude, which favors $\phi\eta$ for the spectroscopic inputs used below. The couplings of the excited vector strangeonia to $K^{(*)}K^{(*)}$ and $\phi\eta^{(\prime)}$ are extracted from the corresponding partial widths:}
\begin{eqnarray}
\label{Eq15}
g_{\phi^*KK}&=&\left[\frac{48 \pi m_{\phi^*}^5 \mathcal{B}(KK)\Gamma_{\phi^*}}{\lambda^{\frac{3}{2}}(m_{\phi^*}^2,m_{K}^2,m_{K}^2)}\right]^{\frac{1}{2}}, \\ \nonumber\\
\label{Eq16} 
g_{\phi^*K^*K}&=&\left[\frac{96 \pi m_{\phi^*}^3 \mathcal{B}(K^*K)\Gamma_{\phi^*}}{\lambda^{\frac{3}{2}}(m_{\phi^*}^2,m_{K^*}^2,m_{K}^2)}\right]^{\frac{1}{2}},\\ \nonumber
\nonumber\\
\nonumber
g_{\phi^*K^*K^*}&=&\left[\frac{192 \pi m_{\phi^*}^5 m_{K^*}^4 \mathcal{B}(K^*K^*)\Gamma_{\phi^*}}{\lambda^{\frac{3}{2}}(m_{\phi^*}^2,m_{K^*}^2,m_{K^*}^2)}\right.\\ \label{Eq17} 
&&\times\left.\frac{1}{(12m_{K^*}^4+20m_{\phi}^2m_{K^*}^2+m_{\phi}^4)}\right]^{\frac{1}{2}},\\ 
\nonumber\\
\label{Eq18}
g_{\phi^*\phi\eta^{(\prime)}}&=&\left[\frac{96 \pi m_{\phi^*}^3 \mathcal{B}(\phi\eta^{(\prime)})\Gamma_{\phi^*}}{\lambda^{\frac{3}{2}}(m_{\phi^*}^2,m_{\phi}^2,m_{\eta^{(\prime)}}^2)}\right]^{\frac{1}{2}},
\end{eqnarray}
{Here, $\lambda(x,y,z)=x^2+y^2+z^2-2xy-2xz-2yz$ is the K\"all\'en function. $\Gamma_{\phi^{*}}$ denotes the total width of the corresponding excited strangeonium state, and the branching fractions are taken from the QPC calculations of Ref.~\cite{Wang:2021gle}, which use the same model parameters.}

\section{The unified description for the cross sections of $e^+e^-\to \phi\eta^{\prime}$ and $ \phi\eta$ }\label{SecIV}
To examine whether the conventional $s\bar{s}$ strangeonium framework can provide a unified description of the measured $e^{+}e^{-}\to\phi\eta$~\cite{BESIII:2021bjn} and $e^{+}e^{-}\to\phi\eta^{\prime}$~\cite{BESIII:2020gnc} line shapes, we perform a simultaneous fit to the two cross-section data sets. {The resonant amplitudes are constructed from the $\phi(2S)$, $\phi(3S)$, and $\phi(2D)$ states. We identify the $\phi(2S)$ with the $\phi(1680)$ and take its mass and total width from the PDG~\cite{ParticleDataGroup:2026aaa}, while the $\phi(3S)$ and $\phi(2D)$ parameters are fixed to those obtained from the open-strange analysis of Ref.~\cite{Wang:2021gle}. The dileptonic widths are also taken from Ref.~\cite{Wang:2021gle}, while the QPC branching fractions for the $\phi\eta$ and $\phi\eta^{\prime}$ decays are calculated here using the same model and parameter set. All three states are included in the $\phi\eta$ amplitude. For the $\phi\eta^{\prime}$ channel, the $\phi(2S)$ lies far below threshold and its off-shell contribution can be absorbed into the continuum background, so only the $\phi(3S)$ and $\phi(2D)$ amplitudes are retained. Once these inputs are fixed, the loop contributions in both channels are controlled by the common cutoff parameter $\alpha_{\Lambda}$. The resonance parameters and decay inputs are summarized in Table~\ref{T1}.}

\begin{table}[!htbp]
  \centering
  \renewcommand\arraystretch{1.5}
  \caption{{Resonance parameters and decay properties of the excited strangeonium states. The mass and total width of the $\phi(2S)$ are taken from the PDG~\cite{ParticleDataGroup:2026aaa}, while those of the $\phi(3S)$ and $\phi(2D)$ and the dileptonic widths are taken from Ref.~\cite{Wang:2021gle}. The $\phi\eta$ and $\phi\eta^{\prime}$ branching fractions are calculated with the QPC model using the same parameter set as Ref.~\cite{Wang:2021gle}.}}\label{T1}
   {\tabcolsep0.11in
  \begin{tabular}{ccccccc}
  \toprule[1pt]
  \midrule[1pt]
  Parameters                        & $\phi(2S)$     & $\phi(3S)$          & $\phi(2D)$ \\
  \midrule[1pt]

  Mass\,(MeV)                       & $1680\pm20$     & $2183\pm1$          & $2290\pm3$\\

  $\Gamma_{\rm{Total}}$\,(MeV)      & $150\pm50$      & $185\pm4$           & $312\pm6$\\

  $\Gamma_{e^+e^-}$\,(eV)           & 285             & 106                 & 17 \\

  $\mathcal{B}(KK)$                 & 0.108           & 0.059               & 0.087\\

  $\mathcal{B}(K^*K)$               & 0.851           & 0.242               & 0.067\\

  $\mathcal{B}(K^*K^*)$             & ...             & 0.007               & 0.105\\

  $\mathcal{B}(\phi\eta)$           & 0.041           & 0.011               & 0.003\\

  $\mathcal{B}(\phi\eta^{\prime})$  & ...             & 0.0003              & 0.0001\\
\midrule[1pt]
\bottomrule[1pt]
\end{tabular}
}
\end{table}

\begin{figure*}[!t]
  \centering
  \begin{tabular}{ccc}
  \subfigure{\label{F11}\includegraphics[width=240pt]{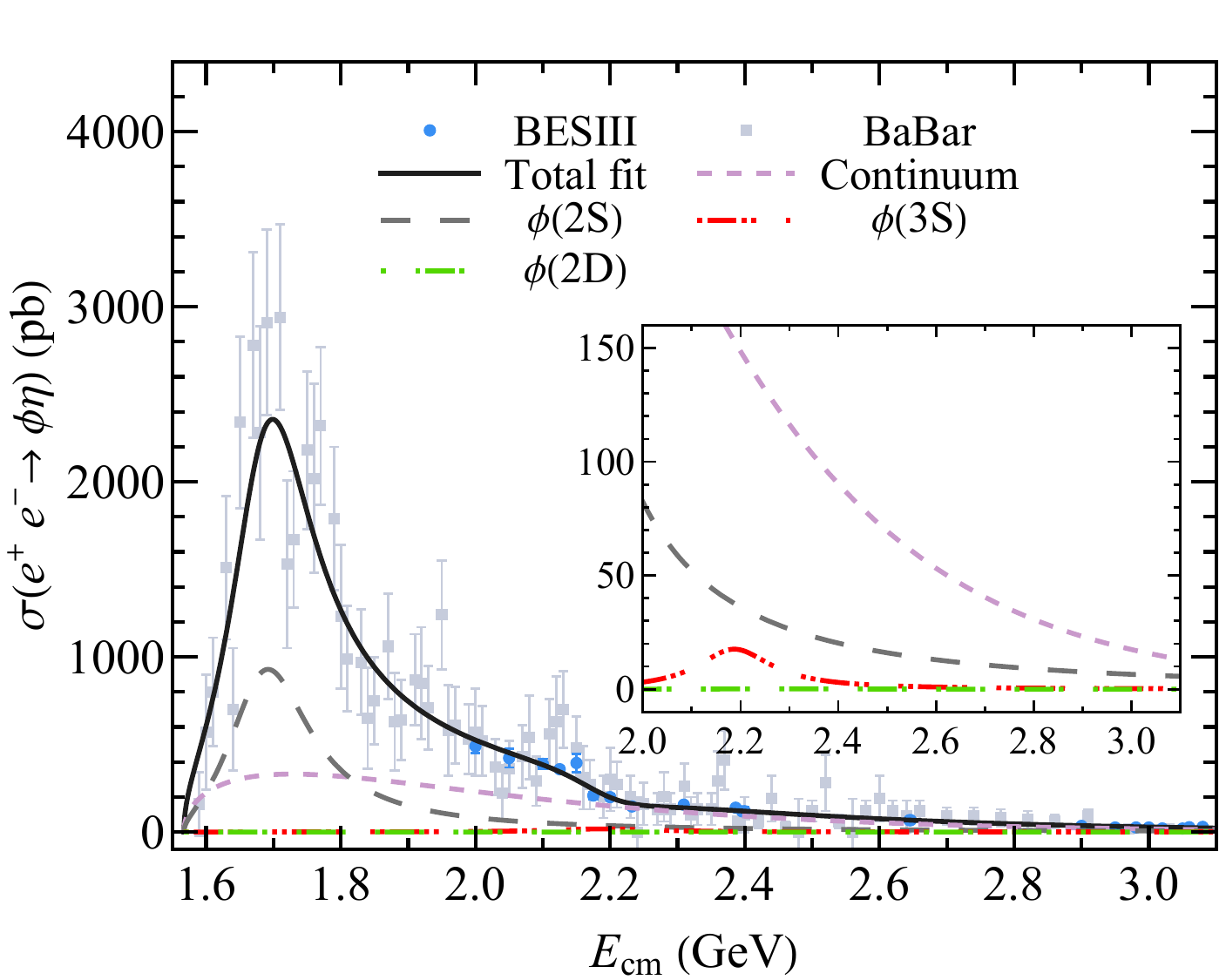}}&$\quad$&\subfigure{\label{F12}\includegraphics[width=240pt]{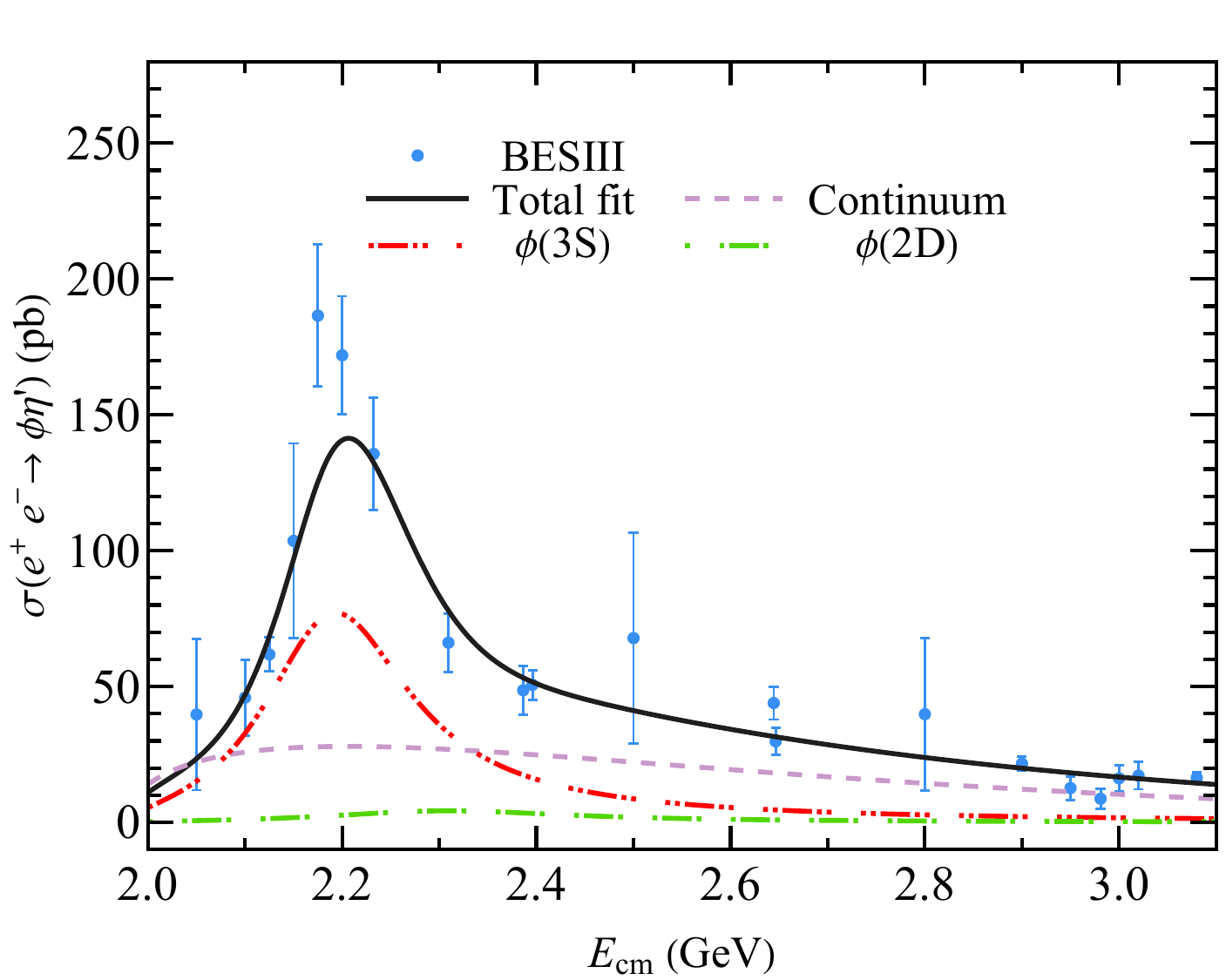}}\\
  (a)&$\quad$&(b)\\
  \end{tabular}
  \caption{{Combined fit to the $e^+e^-\to\phi\eta$~\cite{BaBar:2007ceh,BESIII:2021bjn} and $e^+e^-\to\phi\eta^{\prime}$~\cite{BESIII:2020gnc} cross sections, shown in panels (a) and (b), respectively.}}\label{F2}
\end{figure*}

{The resulting fitted curves are shown in Fig.~\ref{F2}, and the fitted parameters are listed in Table~\ref{T2}. The fit gives $\chi^2/{\rm d.o.f.}=1.023$, indicating that the framework provides a satisfactory overall description of the available $e^+e^-\to\phi\eta$ and $e^+e^-\to\phi\eta'$ data.}
As shown in Fig.~\ref{F2}~(a), the $e^+e^-\to \phi\eta$ cross section at low energies is governed predominantly by the $\phi(2S)$ contribution and the nonresonant continuum. The $\phi(2D)$ and $\phi(3S)$ contribute in the energy region above approximately $2.1~\mathrm{GeV}$. However, the broad $\phi(2D)$ and $\phi(3S)$ contributions do not appear as an obvious structure in the $e^+e^- \to \phi\eta$ cross section.
For the $\phi\eta'$ channel, shown in Fig.~\ref{F2}~(b), the broad enhancement in the $\phi(2170)$ region is reproduced by the coherent interplay of the
$\phi(2D)$, $\phi(3S)$, and continuum amplitudes. The measured structure should therefore not be interpreted simply as the contribution of a single isolated resonance. Instead, it can arise from  interference between the two excited strangeonium states.

\begin{table}[!htbp]
  \centering
  \renewcommand\arraystretch{1.5}
  \caption{Parameters in the combined fit.  The index $i=1$ ($2$)
  denotes the $\phi\eta$ ($\phi\eta^\prime$) channel.  The $a_i$ and $b_i$
  are the continuum parameters in Eq.~\eqref{Eq2}, and
  $\varphi_{ik}$ are the relative phases in Eq.~\eqref{Eq1}.}\label{T2}
   {\tabcolsep0.18in
  \begin{tabular}{cc}
  \toprule[1pt]
  \midrule[1pt]
  Parameters                                &  Values             \\
  \midrule[1pt]
  
  $a_1$\,($\times10^{-3}$)                  &  $2.03\pm0.02$      \\

  $b_1$\,($\mathrm{GeV}^{-1}$)              &  $1.66\pm0.01$      \\

  $\varphi_{11}$\,(\rm{rad})                &  $1.15\pm0.03$      \\

  $\varphi_{12}$\,(\rm{rad})                &  $3.99\pm0.07$      \\

  $\varphi_{13}$\,(rad)                     &  $1.84\pm0.59$      \\
  
  $a_2$\,($\times10^{-3}$)                  &  $0.50\pm0.02$      \\

  $b_2$\,($\mathrm{GeV}^{-1}$)              &  $1.15\pm0.04$      \\

  $\varphi_{22}$\,(\rm{rad})                &  $0.78\pm0.07$      \\
  
  $\varphi_{23}$\,(\rm{rad})                &  $4.64\pm0.18$      \\
  
  $\alpha_{\Lambda}$                        &  $4.23\pm0.24$      \\
  
  $\chi^2/\rm{d.o.f.}$                      &  $1.023$            \\

\midrule[1pt]
\bottomrule[1pt]
\end{tabular}
}
\end{table}

{Table~\ref{T3} lists the loop-induced and total branching fractions of the excited vector strangeonium states to $\phi\eta$ and $\phi\eta'$, evaluated at the fitted value $\alpha_{\Lambda}=4.23\pm0.24$. The loop-induced quantities are obtained from $|\mathcal{M}_{\mathrm{Loop}}|^2$, whereas the total branching fractions are calculated from $|\mathcal{M}_{\mathrm{Tree}}+\mathcal{M}_{\mathrm{Loop}}|^2$. A comparison of Tables~\ref{T1} and~\ref{T3} shows that, for both $\phi(3S)$ and $\phi(2D)$, although the tree-level QPC mechanism strongly favors $\phi\eta$, the strange-meson loop contribution to $\phi\eta^{\prime}$ exceeds that to $\phi\eta$ by approximately one to two orders of magnitude. This enhancement arises primarily from the SU(3)-flavor structure of the $K^{(*)}\bar{K}^{(*)}\eta^{(\prime)}$ vertices, for which the $K^{(*)}\bar{K}^{(*)}\eta^{\prime}$ coupling is about one order of magnitude larger than the $K^{(*)}\bar{K}^{(*)}\eta$ coupling.}
{This coupling hierarchy allows the strange-meson loop contribution to overcome the smaller phase space of the $\phi\eta^{\prime}$ channel.}
{Consequently, after the loop-induced amplitudes are included, the total branching fraction for $\phi\eta^{\prime}$ becomes substantially larger than that for $\phi\eta$, indicating that long-distance strange-meson loop dynamics may be dominant in the hidden-strange decays of excited vector strangeonia.}

\begin{table*}[htb]
\renewcommand\arraystretch{1.5}
\caption{Loop-induced and total branching fractions for the decays of the excited vector strangeonium states into $\phi\eta^{(\prime)}$, evaluated at the fitted value $\alpha_{\Lambda}=4.23\pm0.24$. The loop-induced results are obtained from $|\mathcal{M}_{\rm Loop}|^{2}$, whereas the total results are calculated from $|\mathcal{M}_{\rm Tree}+\mathcal{M}_{\rm Loop}|^{2}$.}
\label{T3}
\begingroup
\setlength{\arrayrulewidth}{1pt}
\begin{ruledtabular}
{\tabcolsep0.18in
\begin{tabular}{lcccccc}
Channels & \multicolumn{2}{c}{$\phi(2S)$}  & \multicolumn{2}{c}{$\phi(3S)$}  &  \multicolumn{2}{c}{$\phi(2D)$}\\

\cline{2-3} \cline{4-5} \cline{6-7}

&$\mathcal{B}_{\rm Loop}\,(\%)$  &  $\mathcal{B}_{\rm Total}\,(\%)$   &   $\mathcal{B}_{\rm Loop}\,(\%)$  &$\mathcal{B}_{\rm Total}\,(\%)$ & $\mathcal{B}_{\rm Loop}\,(\%)$&$\mathcal{B}_{\rm Total}\,(\%)$\\
\hline
$\phi\eta$&$0.79\pm0.12$&$6.59\pm0.18$&$0.13\pm0.02$&$0.99\pm0.01$&$0.05\pm0.01$&$0.10\pm0.01$\\

$\phi\eta^{\prime}$&$\cdots$&$\cdots$&$4.04\pm0.58$&$4.30\pm0.61$&$1.76\pm0.30$&$2.73\pm0.37$
\end{tabular}
}
\end{ruledtabular}
\endgroup
\end{table*}

{We therefore conclude that the enhanced $\phi \eta^{\prime}$ decay mode of the $\phi(2170)$ can be understood by including the $\phi(3S)$ and $\phi(2D)$ states as the relevant intermediate resonances, together with strange-meson loop contributions to their $\phi\eta^{(\prime)}$ transitions. The measured $e^{+}e^{-}\to\phi\eta$ and $e^{+}e^{-}\to\phi\eta^{\prime}$ cross sections can thus be reproduced within the conventional $s\bar{s}$ strangeonium baseline. This interpretation is consistent with the open-strange analysis of Ref.~\cite{Wang:2021gle} and provides a unified description of the open- and hidden-strangeness processes.}

\section{The hints of an additional narrow structure near the $\phi(2170)$}
\label{SecV}

Although the combined fit in Sec.~IV has provided a satisfactory overall description of both cross sections of $e^{+}e^{-}\to\phi\eta$ and $e^{+}e^{-}\to\phi\eta^{\prime}$, it does not reproduce a high-precision BESIII data point near $2.16~\mathrm{GeV}$ shown in Fig.~\ref{F2}~(a). Since the broad structure associated with the $\phi(2170)$ is already described by the excited-strangeonium baseline, this localized deviation motivates us to test an additional, much narrower vector resonance contribution based on the baseline fit in Sec.~IV. We denote this new state by $Y$.
{In Ref.~\cite{Malabarba:2023zez}, the relative decay strengths to $\phi\eta$ and $\phi\eta^{\prime}$ have been proposed as a probe of the internal structure of the $\phi(2170)$ state. Inspired by this work, we fix the ratio $\mathcal{R}_{Y}\equiv\Gamma(Y\to\phi\eta)/\Gamma(Y\to\phi\eta^{\prime})$ to $0.25$, $1$, and $4$ as schemes I, II, and III, respectively, in the following exploratory fit analysis. In each scheme, $M_Y$, $\Gamma_Y$, and $\Gamma^{Y}_{e^{+}e^{-}}{\cal B}(Y\to\phi\eta)$ are free parameters in the combined fit, together with the continuum parameters, relative phases, and common cutoff parameter $\alpha_{\Lambda}$.}

\begin{figure*}[!t]
  \centering
  \begin{tabular}{ccc}
  \subfigure{\label{F11}\includegraphics[width=240pt]{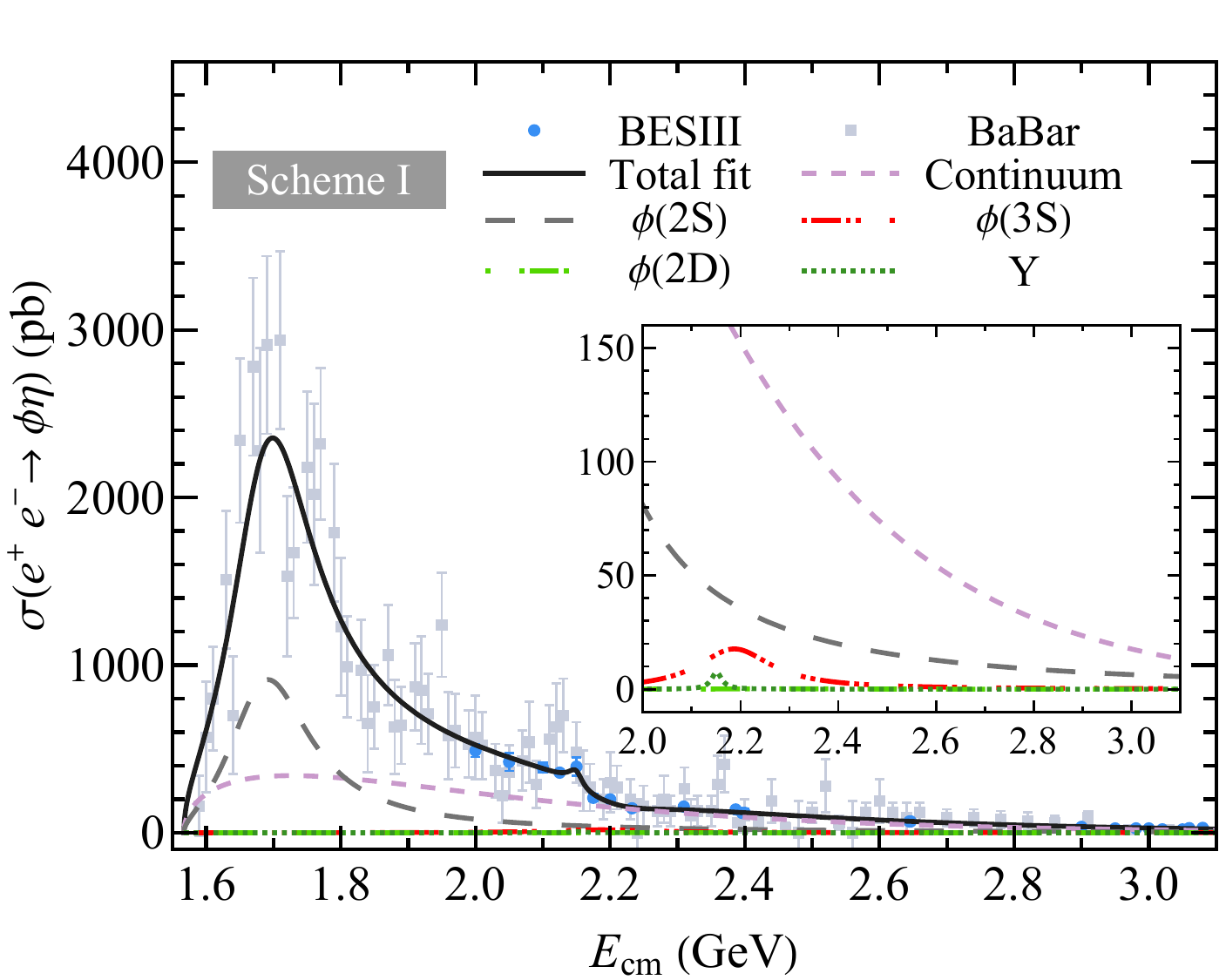}}&$\quad$&\subfigure{\label{F12}\includegraphics[width=240pt]{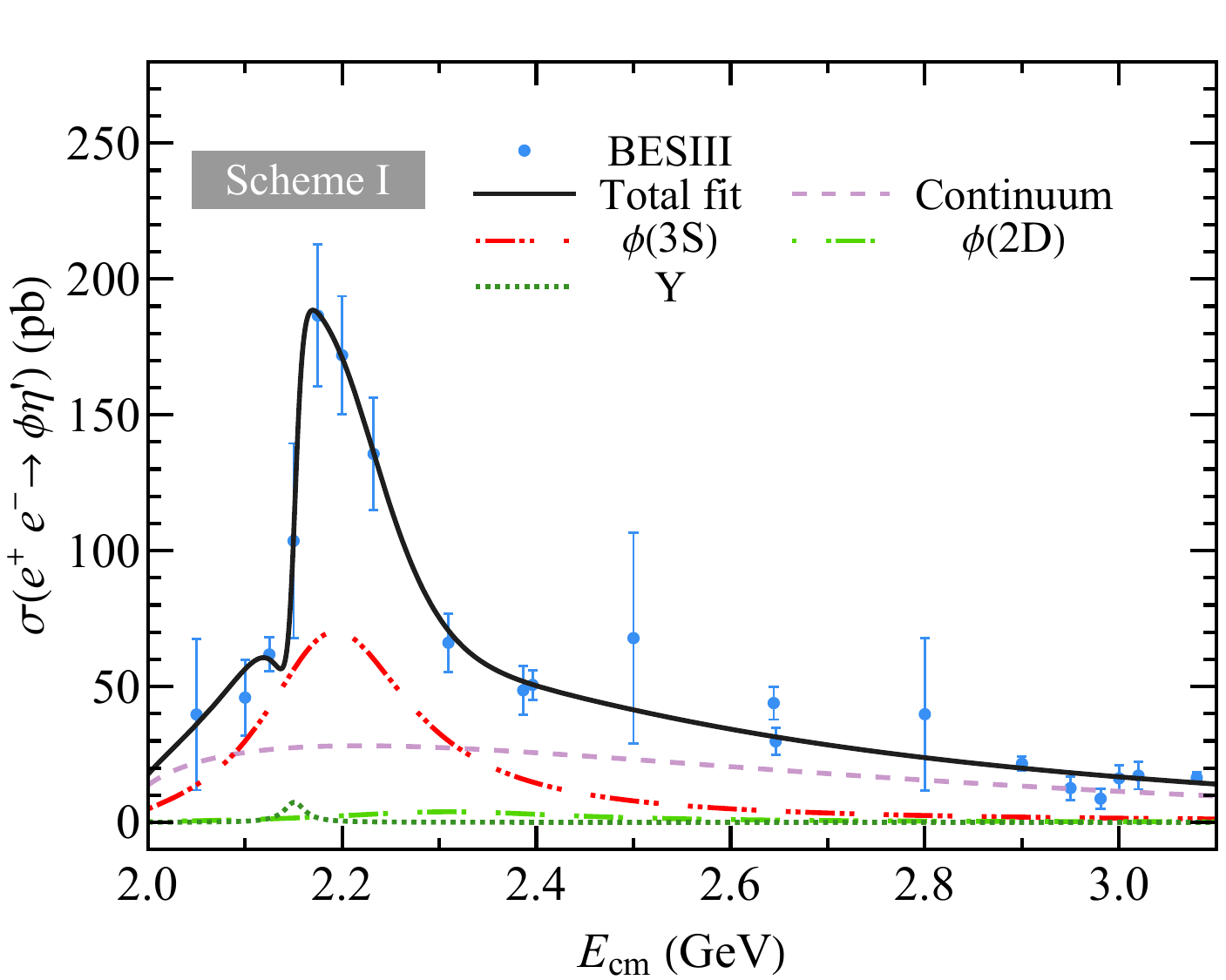}}\\
  (a)&$\quad$&(b)\\
  \subfigure{\label{F11}\includegraphics[width=240pt]{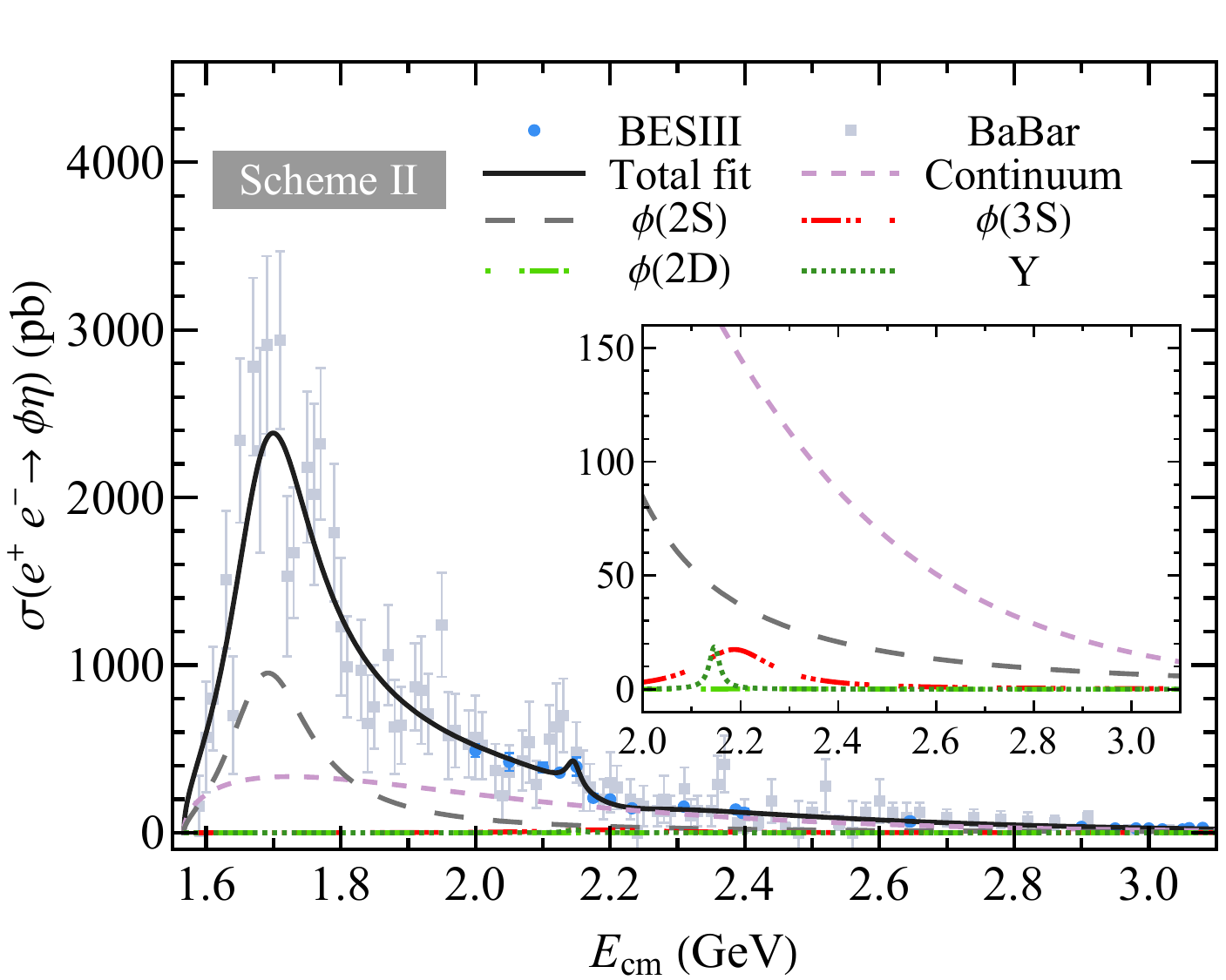}}&$\quad$&\subfigure{\label{F12}\includegraphics[width=240pt]{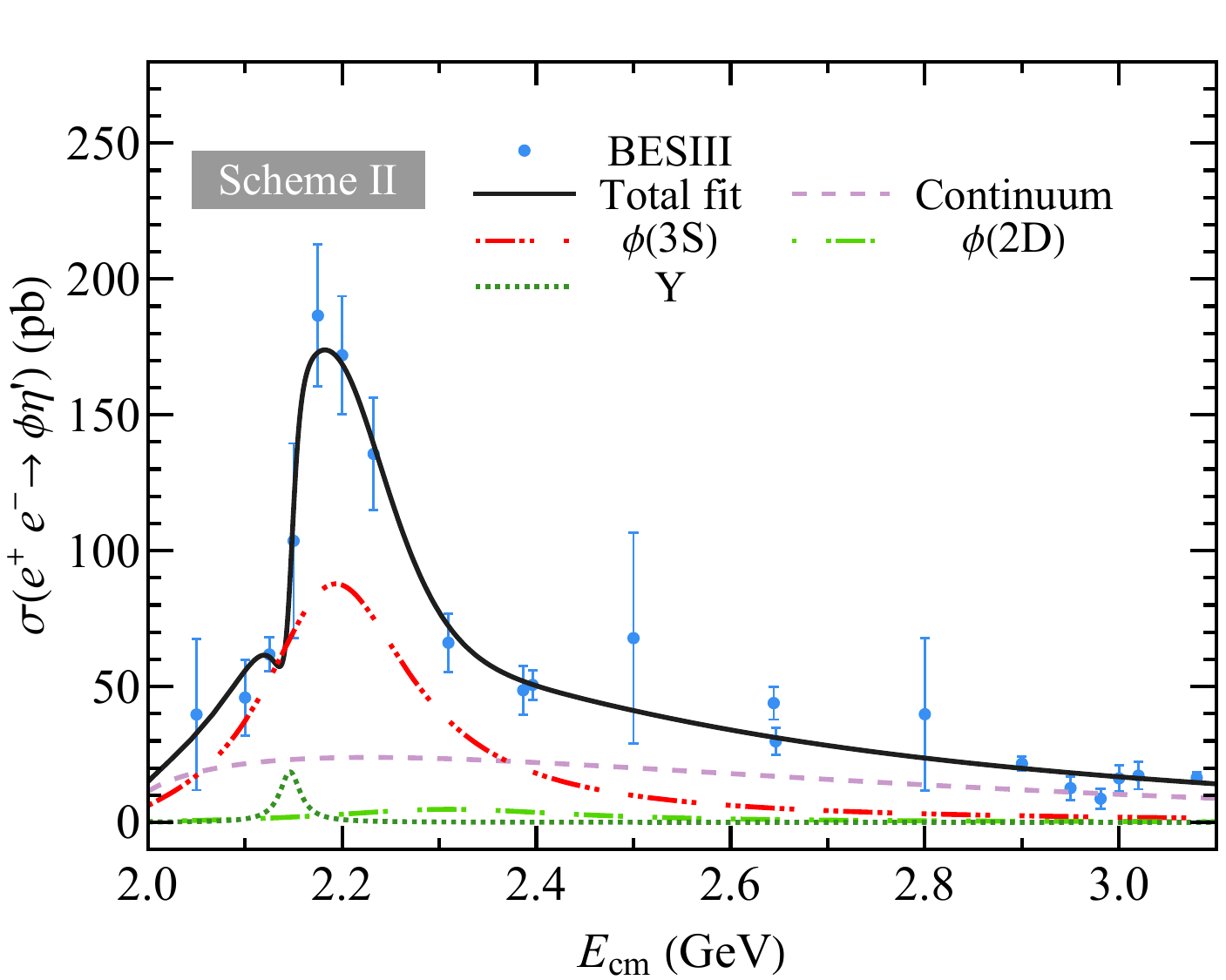}}\\
  (c)&$\quad$&(d)\\
  \subfigure{\label{F11}\includegraphics[width=240pt]{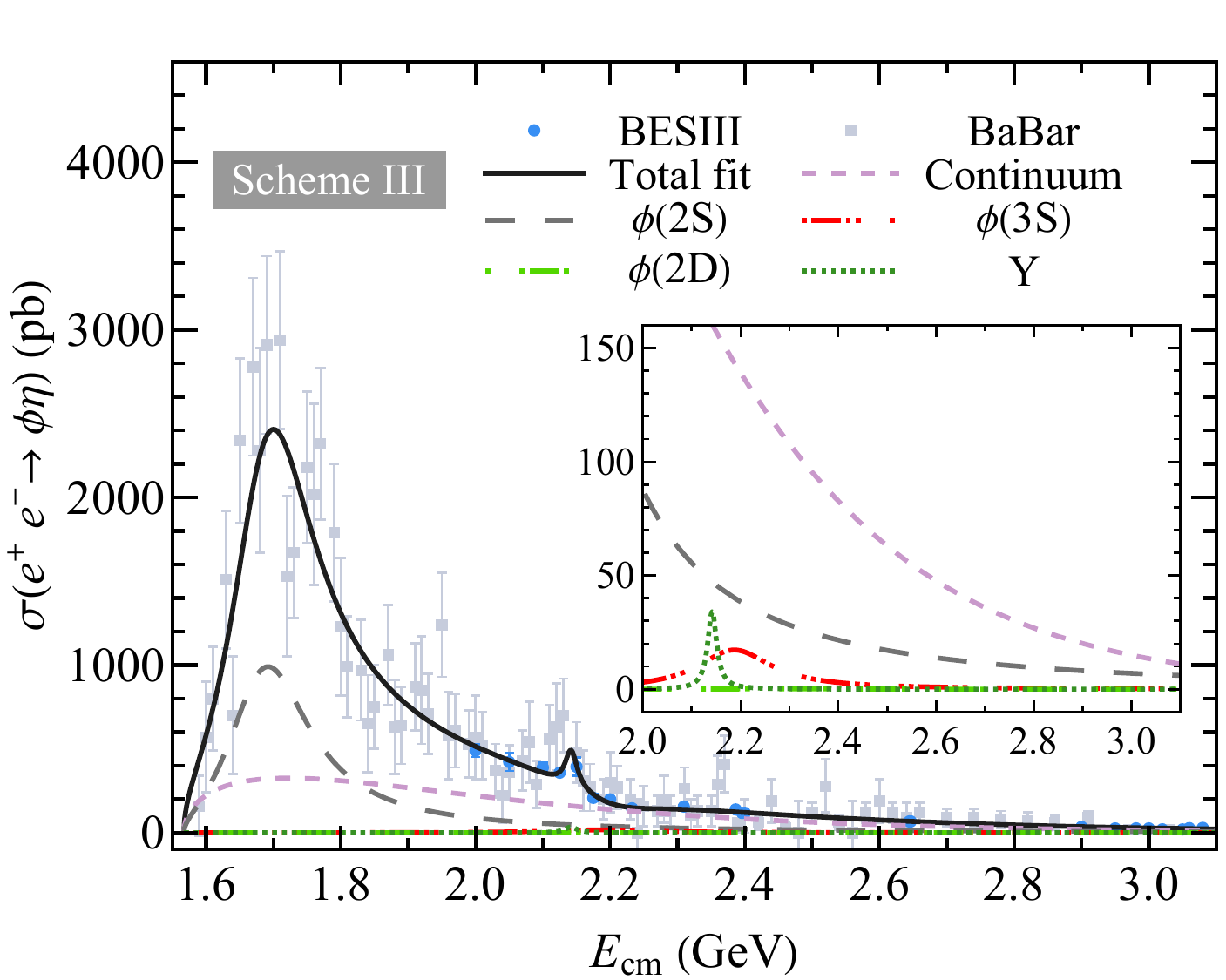}}&$\quad$&\subfigure{\label{F12}\includegraphics[width=240pt]{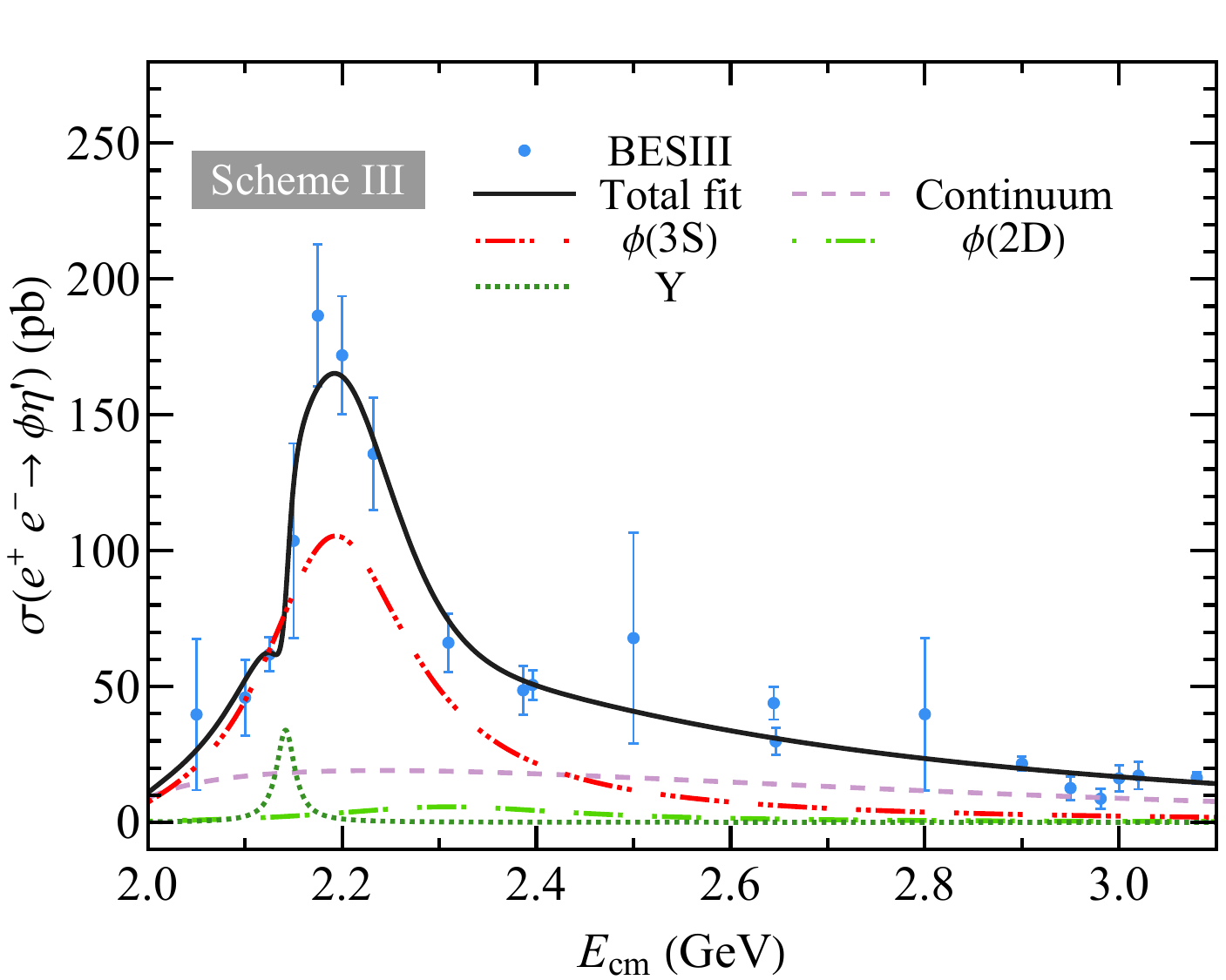}}\\
  (e)&$\quad$&(f)\\
  \end{tabular}
  \caption{{Combined fits to the $e^+e^-\to\phi\eta$~\cite{BaBar:2007ceh,BESIII:2021bjn} and $e^+e^-\to\phi\eta^{\prime}$~\cite{BESIII:2020gnc} cross sections by adding an extra $Y$ state under the three decay-ratio schemes.}}\label{F3}
\end{figure*}



\begin{table*}[htb]
  \centering
  \renewcommand\arraystretch{1.5}
  \caption{Parameters from the combined fits including the additional $Y$
  amplitude. Schemes I, II, and III fix the partial-width ratio $\mathcal{R}_{Y}\equiv\Gamma(Y\to\phi\eta)/\Gamma(Y\to\phi\eta^{\prime})$ to 0.25, 1, and 4, respectively.} \label{T4}

  {\tabcolsep0.18in
  \begin{tabular}{cccc}
  \toprule[1pt]
  \midrule[1pt]
  Parameters  & Scheme I & Scheme II & Scheme III \\
              & $\mathcal{R}_Y=0.25$ & $\mathcal{R}_Y=1$ &
                $\mathcal{R}_Y=4$ \\
  \midrule[1pt]

  $M_{Y}$\,(MeV)                                          &    $2150.69\pm2.87$   &    $2146.25\pm2.73$   &    $2141.67\pm2.10$ \\

  $\Gamma_{Y}$\,(MeV)                                     &    $25.83\pm11.17$    &    $25.72\pm7.30$     &    $22.53\pm5.19$ \\

  $\Gamma_{ee}^{Y}\mathcal{B}(Y\to\phi\eta)$\,(eV)        &    $0.060\pm0.018$    &    $0.150\pm0.040$    &    $0.239\pm0.064$ \\

  $a_1$\,($\times10^{-3}$)                                &    $2.07\pm0.03$      &    $2.06\pm0.03$      &    $2.04\pm0.03$ \\

  $b_1$\,($\mathrm{GeV}^{-1}$)                            &    $1.67\pm0.01$      &    $1.70\pm0.01$      &    $1.72\pm0.01$ \\

  $\varphi_{11}$\,(\rm{rad})                              &    $1.15\pm0.04$      &    $1.10\pm0.04$      &    $1.06\pm0.04$ \\

  $\varphi_{12}$\,(\rm{rad})                              &    $4.18\pm0.07$      &    $4.33\pm0.08$      &    $4.39\pm0.08$ \\

  $\varphi_{13}$\,(\rm{rad})                              &    $2.05\pm0.49$      &    $2.24\pm0.55$      &    $2.35\pm0.59$ \\
  
  $\varphi_{1Y}$\,(rad)                                   &    $2.47\pm0.31$      &    $2.17\pm0.18$      &    $1.96\pm0.13$ \\

  $a_2$\,($\times10^{-3}$)                                &    $0.49\pm0.01$      &    $0.45\pm0.01$      &    $0.40\pm0.01$ \\

  $b_2$\,($\mathrm{GeV}^{-1}$)                            &    $1.10\pm0.04$      &    $1.05\pm0.04$      &    $1.01\pm0.05$ \\

  $\varphi_{22}$\,(\rm{rad})                              &    $1.46\pm0.06$      &    $1.31\pm0.06$      &    $1.07\pm0.07$ \\

  $\varphi_{23}$\,(\rm{rad})                              &    $5.57\pm0.13$      &    $5.45\pm0.12$      &    $5.26\pm0.11$ \\

  $\varphi_{2Y}$\,(rad)                                   &    $5.60\pm0.21$      &    $5.23\pm0.27$      &    $4.94\pm0.41$ \\

  $\alpha_{\Lambda}$                                      &    $4.08\pm0.14$      &    $4.46\pm0.13$      &    $4.81\pm0.13$ \\

  $\chi^2/\rm{d.o.f.}$                                    &    $0.928$            &    $0.926$            &    $0.935$ \\

  \midrule[1pt]
  \bottomrule[1pt]
  \end{tabular}
  }
\end{table*}

The resulting fitted line shapes are shown in Fig.~\ref{F3}, and the fitted parameters are listed in Table~\ref{T4}. The values of $\chi^{2}/\mathrm{d.o.f.}$ and the visual comparison in Fig.~\ref{F3} show that the inclusion of an additional narrow resonance contribution near $2.15~\mathrm{GeV}$ is favored. The fitted $Y$ widths of about $25~\mathrm{MeV}$ can also be compared with the structure reported by BESIII in $e^{+}e^{-}\to K^{*}(892)^{0}\bar K^{0}+\mathrm{c.c.}$, for which $M=(2164.7\pm9.1\pm3.1)~\mathrm{MeV}$ and $\Gamma=(32.4\pm21.0\pm1.8)~\mathrm{MeV}$ were obtained~\cite{BESIII:2023xac}. The widths are consistent within uncertainties, although the fitted $Y$ masses are somewhat lower.

Our analysis indicates a preference for an additional narrow resonance contribution but does not by itself establish a new resonance due to the large bin size of the present measured energy points.   More precise measurements, particularly a finer energy scan in the $2.1$--$2.2~\mathrm{GeV}$ region, together with a combined analysis of the $\phi\eta$, $\phi\eta^{\prime}$, and open-strange final states, will be essential for determining whether an additional vector state is necessary for depicting the observed line shape behavior. If a narrow vector state around 2.15 GeV is eventually confirmed, its small width would make it a particularly interesting candidate for an exotic hadron in the light vector sector. Additionally, the resonance parameters of the additional state and the value of the fitted $\chi^{2}/\mathrm{d.o.f.}$ are relatively insensitive to the assumed values of ${\cal R}_{Y}$. The present data therefore constrain the mass and width of the additional resonance more strongly than its relative couplings to $\phi\eta$ and $\phi\eta^{\prime}$. Consequently, the ratio ${\cal R}_{Y}$ cannot presently distinguish among its potential exotic interpretations such as a hadronic molecule, a compact tetraquark, a strangeonium hybrid, and other exotic configurations.

\section{Summary}\label{SecVI}

The observation of a broad enhancement associated with the $\phi(2170)$ in the 
phase-space-suppressed $e^+e^-\to\phi\eta^\prime$ channel, but not in
$e^+e^-\to\phi\eta$, poses an important puzzle for understanding the
nature of this vector state. We investigate
whether this decay behavior can be understood through the decay
dynamics of excited strangeonia without introducing an exotic
component into the $\phi(2170)$ state. For this purpose, we
adopt the conventional strangeonium baseline established by the
previous analysis of the cross sections for $e^+e^-$ annihilation into open-strange final
states~\cite{Wang:2021gle}, in which the corresponding $\phi(2170)$-like resonant structures are uniformly
described by introducing the interference among $\phi(3S)$, $\phi(2D)$, and nonresonant
amplitudes.

With the resonance parameters and dileptonic widths of the
$\phi(3S)$ and $\phi(2D)$ fixed by the open-strange analysis, we explore
the dynamics governing their decays into $\phi\eta^{(\prime)}$. Both
short-distance QPC transitions and long-distance strange-meson loop
contributions are included in a simultaneous cross-sections analysis of the $e^+e^-$ annihilation to the
$\phi\eta$ and $\phi\eta^\prime$. The short-distance
amplitudes strongly favor $\phi\eta$ and therefore cannot explain the
prominence of the $\phi\eta^\prime$ mode. However, we
find that this phenomenon can be resolved by strange-meson loops. The
SU(3)-flavor structure of the
$K^{(*)}\bar K^{(*)}\eta^{(\prime)}$ vertices strongly enhances the
loop transition to $\phi\eta^\prime$ relative to that to $\phi\eta$,
overcoming the suppressed phase space. The measured cross sections
can consequently be described well overall. This analysis indicates that the broad enhancement associated with $\phi(2170)$ in $e^+e^-\to\phi\eta^\prime$ can still be attributed to the $\phi(3S)$, $\phi(2D)$ contributions and their interference effect as found in the open-strange analysis, which further supports this conventional strangeonium baseline for understanding $\phi(2170)$.

Although this framework reproduces the overall behavior of both the cross
sections of $e^+e^-\to\phi\eta$ and $\phi\eta^\prime$, it does not account for one high-precision BESIII
$\phi\eta$ data point near $2.16~\mathrm{GeV}$. We therefore explore
whether this deviation may indicate an additional narrow
vector contribution, which we denote by $Y$. Since introducing this contribution is only intended
as an exploratory test, we
consider three representative inputs,
$\Gamma(Y\to\phi\eta)/\Gamma(Y\to\phi\eta^\prime)=0.25$, $1$, and $4$,
covering substantially different relative decay strengths. All three
schemes improve the cross section description and favor the inclusion of a narrow resonance with a mass
near $2.15~\mathrm{GeV}$ and a width of about $25~\mathrm{MeV}$, while
giving similarly good fit qualities.  The present results show that its
$\phi\eta/\phi\eta^\prime$ partial width ratio cannot serve as an effective discriminator of its internal structure, because the major
part of the anomalously large $\phi\eta^\prime$ strength has already
been explained by the strange-meson loop dynamics of the broad excited strangeonium
contributions.

Our results highlight the importance of long-distance strange-meson
loops in understanding the hidden-strange decays of excited
strangeonia. Such contributions should be taken into account before
unusual decay patterns are attributed to exotic internal structures.
The possible additional narrow contribution near
$2.15~\mathrm{GeV}$ motivates more precise measurements of the
$e^+e^-\to\phi\eta$ and $e^+e^-\to\phi\eta^\prime$ cross sections in
the future. It also calls for improved
measurements of open-strange channels, particularly
$e^+e^-\to K^{*}(892)^0\bar K^0+\mathrm{c.c.}$, where BESIII reported evidence for 
a similar narrow resonance with a mass near $2.165~\mathrm{GeV}$ and a width
of about $32~\mathrm{MeV}$. If an additional narrow vector state
distinct from the broad $\phi(2170)$ is confirmed, it cannot
be accommodated by the strangeonium baseline
used here and would therefore be a compelling candidate for an exotic hadron in the light vector sector.

\section*{ACKNOWLEDGEMENTS}
Q. S. Zhou is supported by the Natural Science Foundation of Inner Mongolia Autonomous Region (Grant No.2025QN01045), the Research Support Program for High-Level Talents at the Autonomous Region level in the Inner Mongolia Autonomous Region (Grant No. 13100-15112049), the Research Startup Project of Inner Mongolia University (Grant No. 10000-23112101/101), and the National Natural Science Foundation of China (Grant No. 12247101), the Fundamental Research Funds for the Central Universities (Grant No.lzujbky-2025-jdzx07), the Natural Science Foundation of Gansu Province (No.22JR5RA389, No.25JRRA799), and the `111 Center' under Grant No. B20063. 
J. Z. Wang is supported by the National Natural Science Foundation of China under Grants No. 12405088, and No. 12547101, and the Start-up Funds of Chongqing University.
D.~G. is supported by the Yanshan University Research Start-up Fund No. 8190891.

\appendix
\section{{The strange-meson loop amplitudes for $\phi^*\to \phi\eta^{(\prime)}$}}\label{appendix}

\begin{figure*}[!t]
  \centering
  \begin{tabular}{ccccc}
  \subfigure{\label{F1a}\includegraphics[width=140pt]{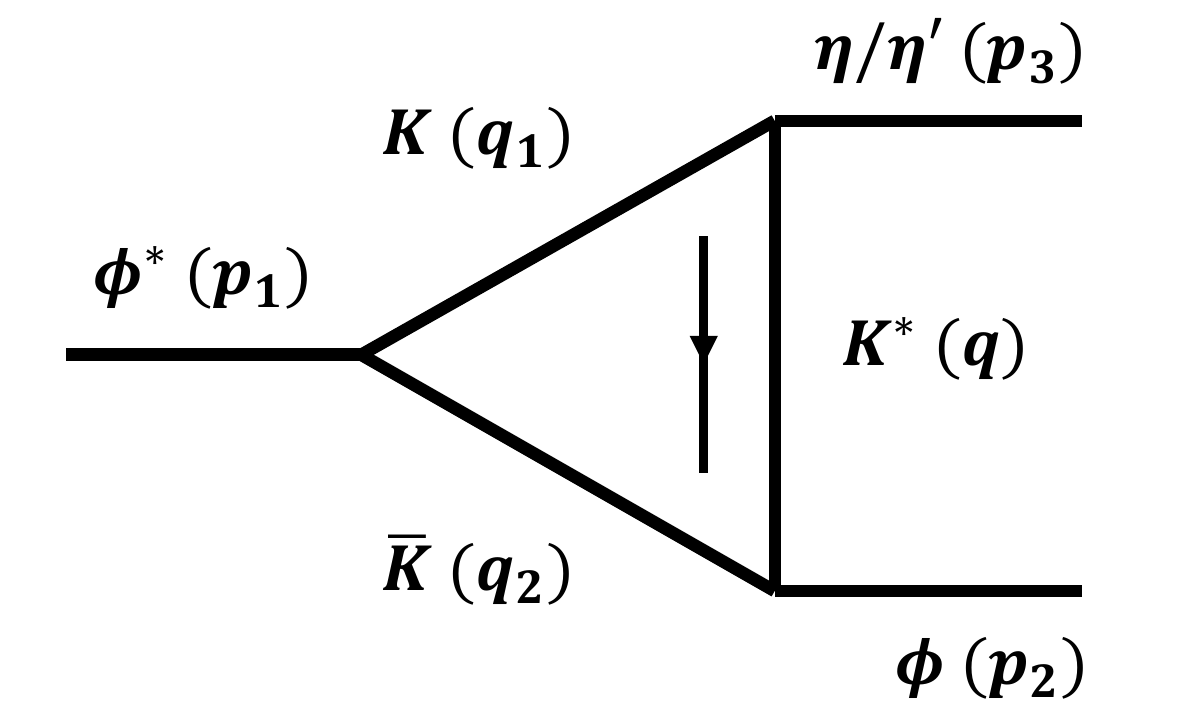}}&$\quad$ &
  \subfigure{\label{F1b}\includegraphics[width=140pt]{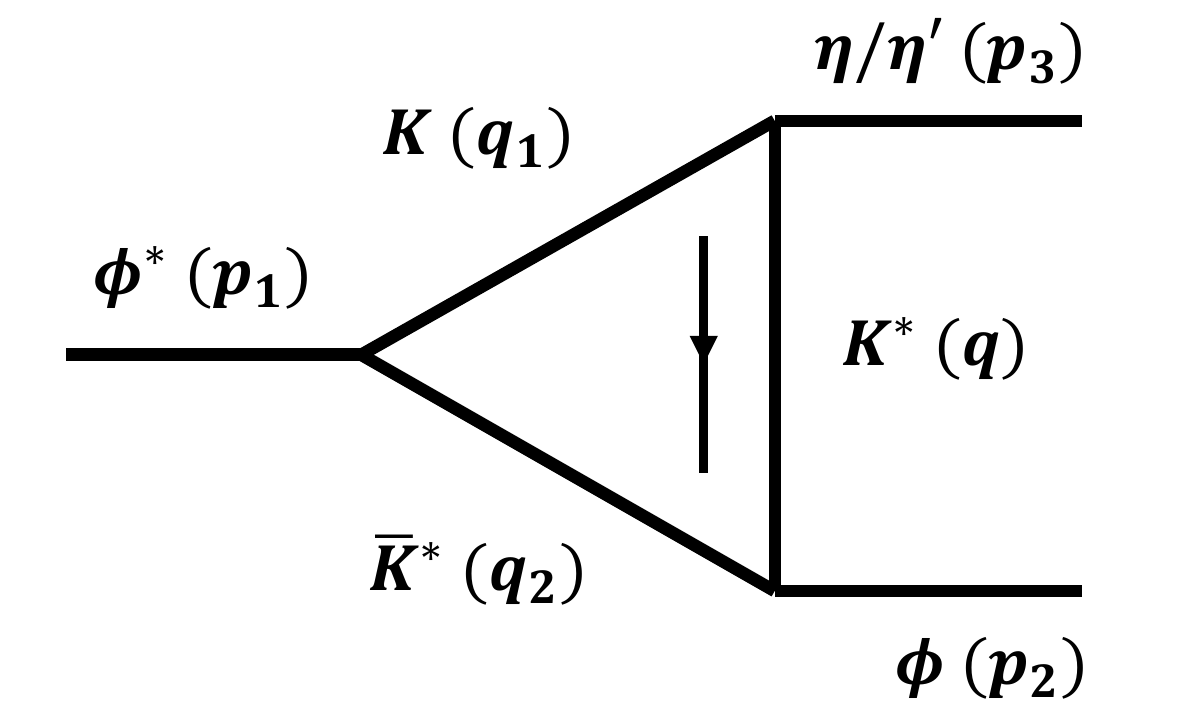}}&$\quad$ &
  \subfigure{\label{F1c}\includegraphics[width=140pt]{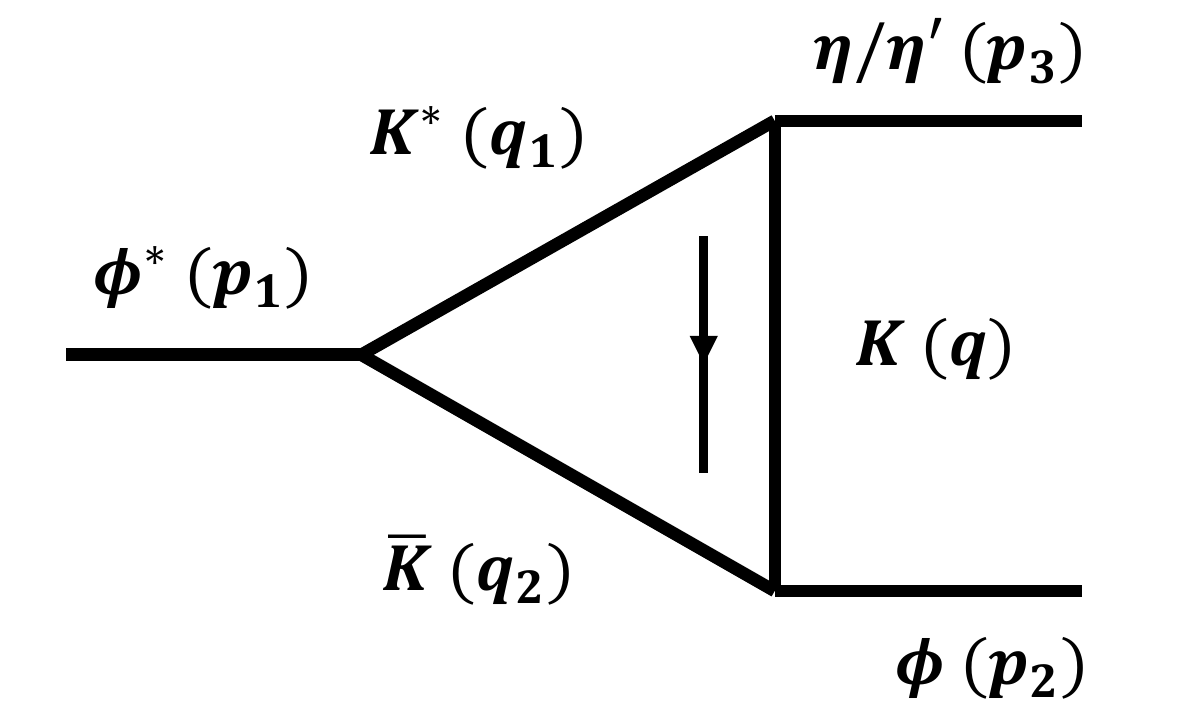}}\\
  (a)&$\quad$&(b)&$\quad$&(c)\\
  \\
  \subfigure{\label{F1d}\includegraphics[width=140pt]{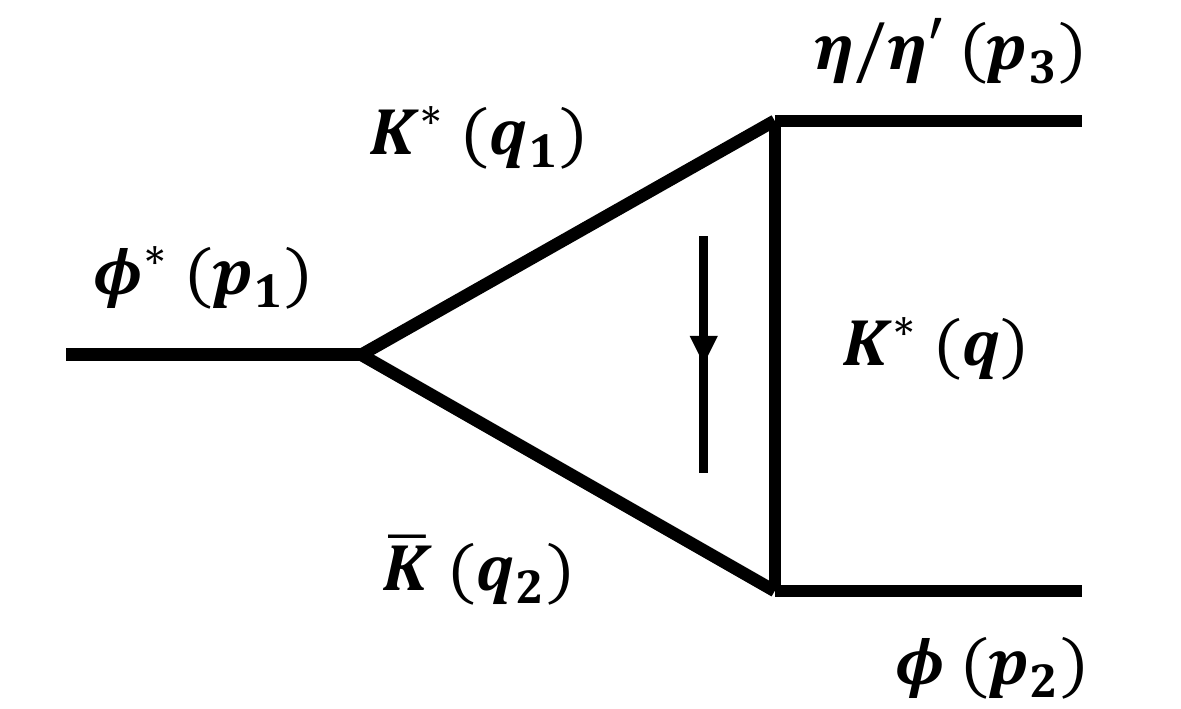}}&$\quad$&
  \subfigure{\label{F1e}\includegraphics[width=140pt]{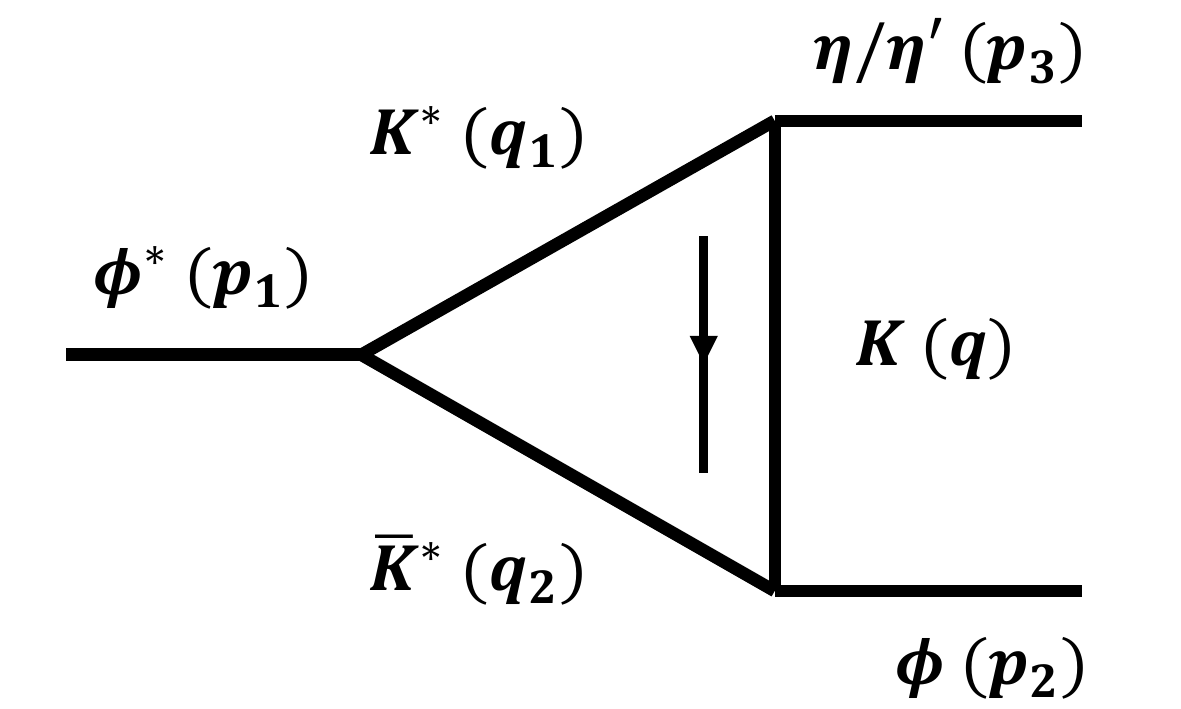}}&$\quad$&
  \subfigure{\label{F1f}\includegraphics[width=140pt]{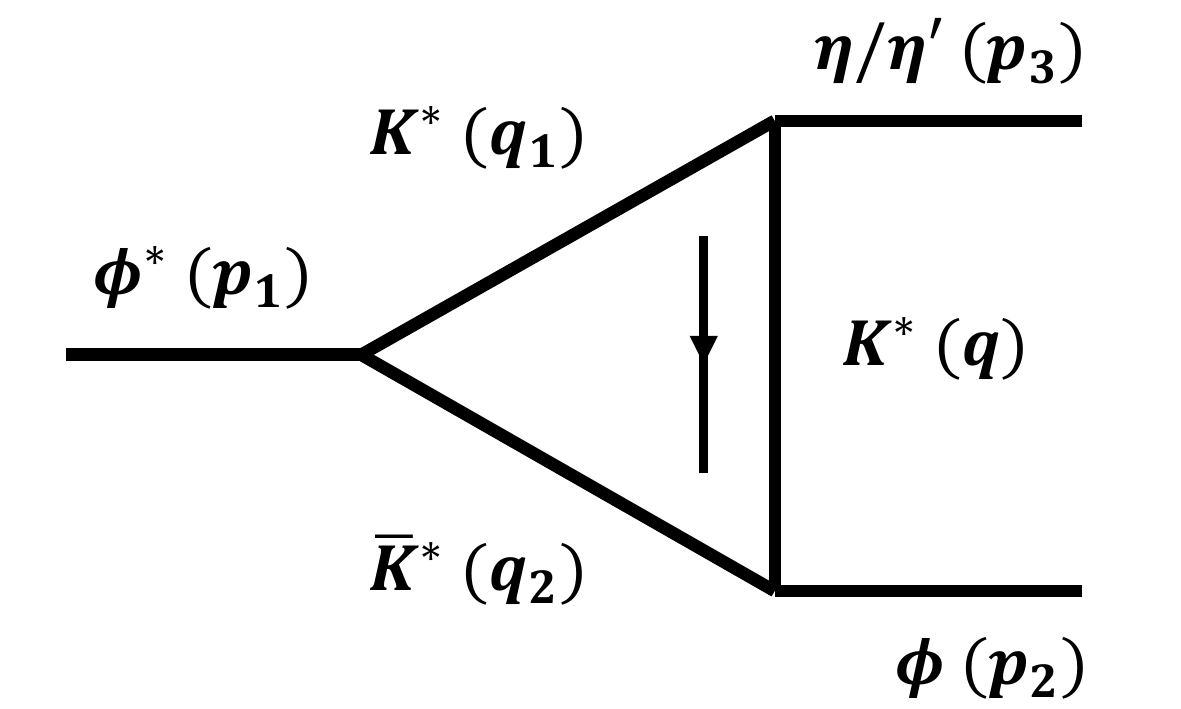}}\\
  (d)&$\quad$&(e)&$\quad$&(f)
  \end{tabular}
  \caption{{Triangle loop topologies contributing to $\phi^*(p_1)\to\phi(p_2)\eta^{(\prime)}(p_3)$. The intermediate mesons and the exchanged meson are (a) $K\bar K$ with $K^*$ exchange, (b) $K\bar K^*$ with $K^*$ exchange, (c) $K^*\bar K$ with $K$ exchange, (d) $K^*\bar K$ with $K^*$ exchange, (e) $K^*\bar K^*$ with $K$ exchange, and (f) $K^*\bar K^*$ with $K^*$ exchange, respectively. The momenta of the intermediate mesons are denoted by $q_1$ and $q_2$, and $q$ is the momentum of the exchanged meson. }}\label{F5}
\end{figure*}

{The loop contribution in Eq.~\eqref{Eq8} is obtained by summing the six triangle diagrams shown in Fig.~\ref{F5}. The initial excited strangeonium $\phi^*(p_1)$ first couples to an allowed $K^{(*)}(q_1)\bar K^{(*)}(q_2)$ pair, which subsequently rescatters into $\phi(p_2)\eta^{(\prime)}(p_3)$ through the exchange of a kaon or vector kaon with momentum $q$. The six diagrams exhaust the combinations of pseudoscalar and vector kaons permitted by the effective Lagrangians in Sec.~\ref{SecIII}. Charge-conjugate and isospin-related loops are included through the overall factor in Eq.~\eqref{Eq8}.}

\begin{widetext}
\begin{eqnarray}
\nonumber
\mathcal{M}^{(1)}_{\mathrm{Loop}}&=&\int\frac{d^4q}{(2\pi)^4}(-ig_{\phi^*KK})  \varepsilon_{\mu}(p_1) (iq_1^{\mu}-iq_2^{\mu})(ig_{K^*K\eta^{(\prime)}})(-iq_1^{\chi}-ip_3^{\chi})
(-g_{\phi K^* K})\epsilon^{\kappa\lambda\theta\delta}(ip_{2\kappa})(-iq_{\theta})\\ \label{eqA1}
&&\times\frac{1}{q_1^2-m_K^2}\frac{1}{q_2^2-m_{K}^2}\frac{\tilde{g}_{\chi\delta}(q)}{q^2-m_{K^*}^2}\varepsilon^{*}_{\lambda}(p_2)\mathcal{F}^2(q,m_{K^*}),\\
\nonumber
\mathcal{M}^{(2)}_{\mathrm{Loop}}&=&\int\frac{d^4q}{(2\pi)^4}(-g_{\phi^*K^*K}) \varepsilon_{\mu}(p_1) \epsilon^{\nu\mu\alpha\beta}(-i p_{1\nu})(i q_{2\alpha})(ig_{K^*K\eta^{(\prime)}})(-iq_1^{\chi}-ip_3^{\chi})(-ig_{\phi K^*K^*})\\ \label{eqA2}
&&\times\left[-ig^{\delta\phi}(q_2^\lambda-q^\lambda)+ig^{\lambda\phi}(p_2^\delta+q_2^\delta)-ig^{\lambda\delta}(q^\phi+p_2^\phi)\right]
\frac{1}{q_1^2-m_K^2}\frac{\tilde{g}_{\beta\phi}(q_2)}{q_2^2-m_{K^*}^2}\frac{\tilde{g}_{\chi\delta}(q)}{q^2-m_{K^*}^2}\varepsilon^{*}_{\lambda}(p_2)\mathcal{F}^2(q,m_{K^*}),\\
\nonumber
\mathcal{M}^{(3)}_{\mathrm{Loop}}&=&\int\frac{d^4q}{(2\pi)^4}
(-g_{\phi^*K^*K}) \varepsilon_{\mu}(p_1) \epsilon^{\nu\mu\rho\sigma}(-i p_{1\nu})(i q_{1\rho})(ig_{K^*K\eta^{(\prime)}})(ip_3^{\iota}-iq^{\iota})(-ig_{\phi KK})(-iq_2^\lambda+iq^\lambda)\\ \label{eqA3}
&&\times\frac{\tilde{g}_{\sigma\iota}(q_1)}{q_1^2-m_{K^*}^2}\frac{1}{q_2^2-m_K^2}\frac{1}{q^2-m_{K}^2}\varepsilon^{*}_{\lambda}(p_2)\mathcal{F}^2(q,m_K),\\
\nonumber
\mathcal{M}^{(4)}_{\mathrm{Loop}}&=&\int\frac{d^4q}{(2\pi)^4}
(-g_{\phi^*K^*K}) \varepsilon_{\mu}(p_1) \epsilon^{\nu\mu\rho\sigma}(-i p_{1\nu})(i q_{1\rho})(g_{K^*K^*\eta^{(\prime)}})\epsilon^{\gamma\iota\varphi\chi}(-iq_{1\gamma})(iq_{\varphi})(-g_{\phi K^*K})\epsilon^{\kappa\lambda\theta\delta}(ip_{2\kappa})(-iq_{\theta})\\ \label{eqA4}
&&\times\frac{\tilde{g}_{\sigma\iota}(q_1)}{q_1^2-m_{K^*}^2}\frac{1}{q_2^2-m_K^2}\frac{\tilde{g}_{\chi\delta}(q)}{q^2-m_{K^*}^2}\varepsilon^{*}_{\lambda}(p_2)\mathcal{F}^2(q,m_{K^*}),\\
\nonumber
\mathcal{M}^{(5)}_{\mathrm{Loop}}&=&\int\frac{d^4q}{(2\pi)^4}
(-ig_{\phi^*K^*K^*}) \varepsilon_{\mu}(p_1) \left[ig^{\sigma\beta}(q_1^{\mu}-q_2^{\mu})-ig^{\mu\sigma}(p_1^{\beta}+q_1^{\beta})+ig^{\mu\beta}(q_2^{\sigma}+p_1^{\sigma})(ig_{K^*K\eta^{(\prime)}})(ip_3^{\iota}-iq^{\iota})\right]\\  \label{eqA5}
&&\times(-g_{\phi K^* K})\epsilon^{\kappa\lambda\eta\phi}(ip_{2\kappa})(-iq_{2\eta})
\frac{\tilde{g}_{\sigma\iota}(q_1)}{q_1^2-m_{K^*}^2}\frac{\tilde{g}_{\beta\phi}(q_2)}{q_2^2-m_{K^*}^2}\frac{1}{q^2-m_{K}^2}\varepsilon^{*}_{\lambda}(p_2)\mathcal{F}^2(q,m_K),\\
\nonumber
\mathcal{M}^{(6)}_{\mathrm{Loop}}&=&\int\frac{d^4q}{(2\pi)^4}
(-ig_{\phi^*K^*K^*}) \varepsilon_{\mu}(p_1) \left[ig^{\sigma\beta}(q_1^{\mu}-q_2^{\mu})-ig^{\mu\sigma}(p_1^{\beta}+q_1^{\beta})+ig^{\mu\beta}(q_2^{\sigma}+p_1^{\sigma})\right](g_{K^*K^*\eta^{(\prime)}})\epsilon^{\gamma\iota\varphi\chi}(-iq_{1\gamma})(iq_{\varphi})(-ig_{\phi K^*K^*})\\ \label{eqA6}
&&\times\left[ -ig^{\delta\phi}(q_2^\lambda-q^\lambda)+ig^{\lambda\phi}(p_2^\delta+q_2^\delta)-ig^{\lambda\delta}(q^\phi+p_2^\phi)\right] \frac{\tilde{g}_{\sigma\iota}(q_1)}{q_1^2-m_{K^*}^2}\frac{\tilde{g}_{\beta\phi}(q_2)}{q_2^2-m_{K^*}^2}\frac{\tilde{g}_{\chi\delta}(q)}{q^2-m_{K^*}^2}\varepsilon^{*}_{\lambda}(p_2)\mathcal{F}^2(q,m_{K^*}).
\end{eqnarray}
\end{widetext}

\bibliography{Ref}

@article{ParticleDataGroup:2026aaa,
    author = "Takahashi, F. and others",
    collaboration = "Particle Data Group",
    title = "{Review of Particle Physics}",
    doi = "10.1142/S0217751X26300115",
    journal = "Int. J. Mod. Phys. A",
    volume = "41",
    pages = "2630011",
    year = "2026"
}

@article{BaBar:2006gsq,
    author = "Aubert, Bernard and others",
    collaboration = "{BABAR Collaboration}",
    title = "{A structure at 2175 MeV in $e^{+}e^{-}\to\phi f_{0}(980)$ observed via initial-state radiation}",
    reportNumber = "SLAC-PUB-12146, BABAR-PUB-06-1482, BABAR-PUB-06-056",
    doi = "10.1103/PhysRevD.74.091103",
    journal = "Phys. Rev. D",
    volume = "74",
    pages = "091103",
    year = "2006"
}

@article{Belle:2008kuo,
    author = "Shen, C. P. and others",
    collaboration = "{Belle Collaboration}",
    title = "{Observation of the $\phi(1680)$ and the $Y(2175)$ in $e^{+}e^{-}\to\phi\pi^{+}\pi^{-}$}",
    reportNumber = "BELLE-CONF-0865",
    doi = "10.1103/PhysRevD.80.031101",
    journal = "Phys. Rev. D",
    volume = "80",
    pages = "031101",
    year = "2009"
}

@article{BaBar:2011btv,
    author = "Lees, J. P. and others",
    collaboration = "{BABAR Collaboration}",
    title = "{Cross sections for the reactions $e^{+}e^{-}\to K^{+}K^{-}\pi^{+}\pi^{-}$, $K^{+}K^{-}\pi^{0}\pi^{0}$, and $K^{+}K^{-}K^{+}K^{-}$ measured using initial-state radiation events}",
    reportNumber = "SLAC-PUB-14403, BABAR-PUB-11-001",
    doi = "10.1103/PhysRevD.86.012008",
    journal = "Phys. Rev. D",
    volume = "86",
    pages = "012008",
    year = "2012"
}

@article{BESIII:2021aet,
    author = "Ablikim, Medina and others",
    collaboration = "{BESIII Collaboration}",
    title = "{Measurement of $e^{+}e^{-}\to\phi\pi^{+}\pi^{-}$ cross sections at center-of-mass energies from 2.00 to 3.08~GeV}",
    doi = "10.1103/PhysRevD.108.032011",
    journal = "Phys. Rev. D",
    volume = "108",
    number = "3",
    pages = "032011",
    year = "2023"
}

@article{BaBar:2007ceh,
    author = "Aubert, Bernard and others",
    collaboration = "{BABAR Collaboration}",
    title = "{Measurements of $e^{+}e^{-}\to K^{+}K^{-}\eta$, $K^{+}K^{-}\pi^{0}$, and $K_S^0K^{\pm}\pi^{\mp}$ cross sections using initial-state radiation events}",
    reportNumber = "SLAC-PUB-12968, BABAR-PUB-07-052",
    doi = "10.1103/PhysRevD.77.092002",
    journal = "Phys. Rev. D",
    volume = "77",
    pages = "092002",
    year = "2008"
}

@article{BESIII:2021bjn,
    author = "Ablikim, Medina and others",
    collaboration = "{BESIII Collaboration}",
    title = "{Study of the process $e^{+}e^{-}\to\phi\eta$ at center-of-mass energies between 2.00 and 3.08~GeV}",
    doi = "10.1103/PhysRevD.104.032007",
    journal = "Phys. Rev. D",
    volume = "104",
    number = "3",
    pages = "032007",
    year = "2021"
}

@article{Belle:2022fhh,
    author = "Zhu, W. J. and others",
    collaboration = "{Belle Collaboration}",
    title = "{Study of $e^{+}e^{-}\to\eta\phi$ via initial-state radiation at Belle}",
    doi = "10.1103/PhysRevD.107.012006",
    journal = "Phys. Rev. D",
    volume = "107",
    pages = "012006",
    year = "2023"
}

@article{BESIII:2020gnc,
    author = "Ablikim, M. and others",
    collaboration = "{BESIII Collaboration}",
    title = "{Observation of a structure in $e^{+}e^{-}\to\phi\eta^{\prime}$ at $\sqrt{s}$ from 2.05 to 3.08~GeV}",
    doi = "10.1103/PhysRevD.102.012008",
    journal = "Phys. Rev. D",
    volume = "102",
    number = "1",
    pages = "012008",
    year = "2020"
}

@article{BESIII:2018ldc,
    author = "Ablikim, M. and others",
    collaboration = "{BESIII Collaboration}",
    title = "{Measurement of the $e^{+}e^{-}\to K^{+}K^{-}$ cross section at $\sqrt{s}=2.00$--$3.08~\mathrm{GeV}$}",
    doi = "10.1103/PhysRevD.99.032001",
    journal = "Phys. Rev. D",
    volume = "99",
    number = "3",
    pages = "032001",
    year = "2019"
}

@article{BESIII:2020vtu,
    author = "Ablikim, M. and others",
    collaboration = "{BESIII Collaboration}",
    title = "{Observation of a resonant structure in $e^{+}e^{-}\to K^{+}K^{-}\pi^{0}\pi^{0}$}",
    doi = "10.1103/PhysRevLett.124.112001",
    journal = "Phys. Rev. Lett.",
    volume = "124",
    number = "11",
    pages = "112001",
    year = "2020"
}

@article{BESIII:2021yam,
    author = "Ablikim, Medina and others",
    collaboration = "{BESIII Collaboration}",
    title = "{Cross-section measurement of $e^{+}e^{-}\to K_S^0K_L^0$ at $\sqrt{s}=2.00$--$3.08~\mathrm{GeV}$}",
    doi = "10.1103/PhysRevD.104.092014",
    journal = "Phys. Rev. D",
    volume = "104",
    number = "9",
    pages = "092014",
    year = "2021"
}

@article{BESIII:2022wxz,
    author = "Ablikim, M. and others",
    collaboration = "{BESIII Collaboration}",
    title = "{Measurement of the $e^{+}e^{-}\to K^{+}K^{-}\pi^{0}$ cross section and observation of a resonant structure}",
    doi = "10.1007/JHEP07(2022)045",
    journal = "J. High Energy Phys.",
    volume = "07",
    number = "2022",
    pages = "045",
    year = "2022"
}

@article{BESIII:2023xac,
    author = "Ablikim, Medina and others",
    collaboration = "{BESIII Collaboration}",
    title = "{Measurement of the $e^{+}e^{-}\to K_S^0K_L^0\pi^0$ cross sections from $\sqrt{s}=2.000$ to $3.080~\mathrm{GeV}$}",
    doi = "10.1007/JHEP01(2024)180",
    journal = "J. High Energy Phys.",
    volume = "01",
    number = "2024",
    pages = "180",
    year = "2024"
}

@article{BaBar:2013jqz,
    author = "Lees, J. P. and others",
    collaboration = "{BABAR Collaboration}",
    title = "{Precision measurement of the $e^{+}e^{-}\to K^{+}K^{-}(\gamma)$ cross section with the initial-state radiation method at {BABAR}}",
    reportNumber = "BABAR-PUB-13-006, SLAC-PUB-15487",
    doi = "10.1103/PhysRevD.88.032013",
    journal = "Phys. Rev. D",
    volume = "88",
    number = "3",
    pages = "032013",
    year = "2013"
}

@article{BaBar:2022ahi,
    author = "Lees, J. P. and others",
    collaboration = "{BABAR Collaboration}",
    title = "{Study of the reactions $e^{+}e^{-}\to K^{+}K^{-}\pi^{0}\pi^{0}\pi^{0}$, $e^{+}e^{-}\to K_S^0K^{\pm}\pi^{\mp}\pi^{0}\pi^{0}$, and $e^{+}e^{-}\to K_S^0K^{\pm}\pi^{\mp}\pi^{+}\pi^{-}$ at center-of-mass energies from threshold to 4.5~GeV using initial-state radiation}",
    reportNumber = "SLAC-PUB-17694",
    doi = "10.1103/PhysRevD.107.072001",
    journal = "Phys. Rev. D",
    volume = "107",
    number = "7",
    pages = "072001",
    year = "2023"
}

@article{Wang:2021gle,
    author = "Wang, Jun Zhang and Wang, Li Ming and Liu, Xiang and Matsuki, Takayuki",
    title = "{Deciphering the light vector meson contribution to the cross sections of $e^{+}e^{-}$ annihilations into the open-strange channels through a combined analysis}",
    doi = "10.1103/PhysRevD.104.054045",
    journal = "Phys. Rev. D",
    volume = "104",
    number = "5",
    pages = "054045",
    year = "2021"
}

@article{Ding:2007pc,
    author = "Ding, Gui Jun and Yan, Mu Lin",
    title = "{$Y(2175)$: Distinguish hybrid state from higher quarkonium}",
    doi = "10.1016/j.physletb.2007.10.020",
    journal = "Phys. Lett. B",
    volume = "657",
    pages = "49--54",
    year = "2007"
}

@article{Wang:2012wa,
    author = "Wang, Xiao and Sun, Zhi Feng and Chen, Dian Yong and Liu, Xiang and Matsuki, Takayuki",
    title = "{Nonstrange partner of strangeonium-like state $Y(2175)$}",
    doi = "10.1103/PhysRevD.85.074024",
    journal = "Phys. Rev. D",
    volume = "85",
    pages = "074024",
    year = "2012"
}

@article{Pang:2019ttv,
    author = "Pang, Cheng Qun",
    title = "{Excited states of $\phi$ meson}",
    doi = "10.1103/PhysRevD.99.074015",
    journal = "Phys. Rev. D",
    volume = "99",
    number = "7",
    pages = "074015",
    year = "2019"
}

@article{Li:2020xzs,
    author = {Li, Qi and Gui, Long Cheng and Liu, Ming Sheng and L{\"u}, Qi Fang and Zhong, Xian Hui},
    title = "{Mass spectrum and strong decays of strangeonium in a constituent quark model}",
    doi = "10.1088/1674-1137/abcf22",
    journal = "Chin. Phys. C",
    volume = "45",
    number = "2",
    pages = "023116",
    year = "2021"
}

@article{Badalian:2019xir,
    author = "Badalian, A. M. and Bakker, B. L. G.",
    title = "{The Regge trajectories and leptonic widths of the vector $s\bar s$ mesons}",
    doi = "10.1007/s00601-019-1525-9",
    journal = "Few Body Syst.",
    volume = "60",
    number = "3",
    pages = "58",
    year = "2019"
}

@article{Feng:2021igh,
    author = {Feng, Jie Cheng and Kang, Xian Wei and L{\"u}, Qi Fang and Zhang, Feng Shou},
    title = "{Possible assignment of excited light $^{3}S_{1}$ vector mesons}",
    doi = "10.1103/PhysRevD.104.054027",
    journal = "Phys. Rev. D",
    volume = "104",
    number = "5",
    pages = "054027",
    year = "2021"
}

@article{Hao:2024nvx,
    author = "Hao, Wei and Sultan, M. Atif and Liu, Li Juan and Wang, En",
    title = "{Strangeonium spectrum with screening effects and interpretation of $h_{1}(1911)$ and $X(2300)$ observed by BESIII}",
    doi = "10.1103/yqxf-5scf",
    journal = "Phys. Rev. D",
    volume = "112",
    pages = "036006",
    year = "2025"
}

@article{Ding:2006ya,
    author = "Ding, Gui Jun and Yan, Mu Lin",
    title = "{A candidate for $1^{--}$ strangeonium hybrid}",
    reportNumber = "USTC-ICTS-06-14",
    doi = "10.1016/j.physletb.2007.05.026",
    journal = "Phys. Lett. B",
    volume = "650",
    pages = "390--400",
    year = "2007"
}

@article{Ho:2019org,
    author = "Ho, J. and Berg, R. and Steele, T. G. and Chen, W. and Harnett, D.",
    title = "{Is the $Y(2175)$ a strangeonium hybrid meson?}",
    doi = "10.1103/PhysRevD.100.034012",
    journal = "Phys. Rev. D",
    volume = "100",
    number = "3",
    pages = "034012",
    year = "2019"
}

@article{Ma:2020bex,
    author = "Ma, Yunheng and Chen, Ying and Gong, Ming and Liu, Zhaofeng",
    title = "{Strangeonium-like hybrids on the lattice}",
    doi = "10.1088/1674-1137/abc241",
    journal = "Chin. Phys. C",
    volume = "45",
    number = "1",
    pages = "013112",
    year = "2021"
}

@misc{Guo:2007uz,
    author = "Guo, Feng Kun and Shen, Peng Nian and Wang, Zhi Gang and Liang, Wei Hong and Kisslinger, L. S.",
    title = "{Light vector hybrid states via QCD sum rules}",
    eprint = "hep-ph/0703062",
    archivePrefix = "arXiv",
    year = "2007"
}

@article{Li:2025hsp,
    author = "Li, Shuang Hong and Huang, Zhuo Ran and Chen, Wei and Jin, Hong Ying",
    title = "{Revising the mass of light hybrid mesons: NLO QCD sum rules point to $\phi(2170)$ as a prime candidate}",
    doi = "10.1007/JHEP03(2026)087",
    journal = "J. High Energy Phys.",
    volume = "03",
    number = "2026",
    pages = "087",
    year = "2026"
}

@article{Wang:2006ri,
    author = "Wang, Zhi Gang",
    title = "{Analysis of $Y(2175)$ as a tetraquark state with QCD sum rules}",
    doi = "10.1016/j.nuclphysa.2007.04.012",
    journal = "Nucl. Phys. A",
    volume = "791",
    pages = "106--116",
    year = "2007"
}

@article{Chen:2008ej,
    author = "Chen, Hua Xing and Liu, Xiang and Hosaka, Atsushi and Zhu, Shi Lin",
    title = "{$Y(2175)$ state in the QCD sum rule}",
    doi = "10.1103/PhysRevD.78.034012",
    journal = "Phys. Rev. D",
    volume = "78",
    pages = "034012",
    year = "2008"
}

@article{Drenska:2008gr,
    author = "Drenska, N. V. and Faccini, R. and Polosa, A. D.",
    title = "{Higher tetraquark particles}",
    doi = "10.1016/j.physletb.2008.09.038",
    journal = "Phys. Lett. B",
    volume = "669",
    pages = "160--166",
    year = "2008"
}

@article{Jiang:2023atq,
    author = "Jiang, Yi Wei and Tan, Wei Han and Chen, Hua Xing and Cui, Er Liang",
    title = "{Strong decays of the $\phi(2170)$ as a fully strange tetraquark state}",
    doi = "10.3390/sym16081021",
    journal = "Symmetry",
    volume = "16",
    number = "8",
    pages = "1021",
    year = "2024"
}

@article{Deng:2010zzd,
    author = "Deng, Chengrong and Ping, Jialun and Wang, Fan and Goldman, T.",
    title = "{Tetraquark state and multibody interaction}",
    doi = "10.1103/PhysRevD.82.074001",
    journal = "Phys. Rev. D",
    volume = "82",
    pages = "074001",
    year = "2010"
}

@article{Chen:2018kuu,
    author = "Chen, Hua Xing and Shen, Cheng Ping and Zhu, Shi Lin",
    title = "{A possible partner state of the $Y(2175)$}",
    doi = "10.1103/PhysRevD.98.014011",
    journal = "Phys. Rev. D",
    volume = "98",
    number = "1",
    pages = "014011",
    year = "2018"
}

@article{Ke:2018evd,
    author = "Ke, Hong Wei and Li, Xue Qian",
    title = "{Study of the strong decays of $\phi(2170)$ and the future charm-tau factory}",
    doi = "10.1103/PhysRevD.99.036014",
    journal = "Phys. Rev. D",
    volume = "99",
    number = "3",
    pages = "036014",
    year = "2019"
}

@article{Agaev:2019coa,
    author = "Agaev, S. S. and Azizi, K. and Sundu, H.",
    title = "{Nature of the vector resonance $Y(2175)$}",
    doi = "10.1103/PhysRevD.101.074012",
    journal = "Phys. Rev. D",
    volume = "101",
    number = "7",
    pages = "074012",
    year = "2020"
}

@article{Liu:2020lpw,
    author = "Liu, Feng Xiao and Liu, Ming Sheng and Zhong, Xian Hui and Zhao, Qiang",
    title = "{Fully strange tetraquark $ss\bar{s}\bar{s}$ spectrum and possible experimental evidence}",
    doi = "10.1103/PhysRevD.103.016016",
    journal = "Phys. Rev. D",
    volume = "103",
    number = "1",
    pages = "016016",
    year = "2021"
}

@article{Su:2022eun,
    author = "Su, Niu and Chen, Hua Xing",
    title = "{S- and P-wave fully strange tetraquark states from QCD sum rules}",
    doi = "10.1103/PhysRevD.106.014023",
    journal = "Phys. Rev. D",
    volume = "106",
    number = "1",
    pages = "014023",
    year = "2022"
}

@article{Xin:2022qnv,
    author = "Xin, Qi and Wang, Zhi Gang",
    title = "{Fully-light vector tetraquark states with explicit $P$-wave via QCD sum rules}",
    doi = "10.1088/1674-1137/ad181c",
    journal = "Chin. Phys. C",
    volume = "48",
    number = "3",
    pages = "033104",
    year = "2024"
}

@article{MartinezTorres:2008gy,
    author = "Mart{\'i}nez Torres, A. and Khemchandani, K. P. and Geng, L. S. and Napsuciale, M. and Oset, E.",
    title = "{$X(2175)$ as a resonant state of the $\phi K\bar{K}$ system}",
    doi = "10.1103/PhysRevD.78.074031",
    journal = "Phys. Rev. D",
    volume = "78",
    pages = "074031",
    year = "2008"
}

@article{Malabarba:2023zez,
    author = "Malabarba, Brenda B. and Khemchandani, K. P. and Mart{\'i}nez Torres, A.",
    title = "{$\phi(2170)$ decaying to $\phi\eta$ and $\phi\eta^{\prime}$}",
    doi = "10.1103/PhysRevD.108.036010",
    journal = "Phys. Rev. D",
    volume = "108",
    number = "3",
    pages = "036010",
    year = "2023"
}

@article{Alvarez-Ruso:2009vkn,
    author = "Alvarez-Ruso, L. and Oller, J. A. and Alarc{\'o}n, J. M.",
    title = "{On the $\phi(1020)f_{0}(980)$ $S$-wave scattering and the $Y(2175)$ resonance}",
    doi = "10.1103/PhysRevD.80.054011",
    journal = "Phys. Rev. D",
    volume = "80",
    pages = "054011",
    year = "2009"
}

@article{Oller:2010tr,
    author = "Oller, J. A. and Alarc{\'o}n, J. M. and Albaladejo, M. and Alvarez-Ruso, L. and Roca, L.",
    title = "{Hadron resonances generated from the dynamics of the lightest scalar ones}",
    doi = "10.1016/j.nuclphysbps.2010.10.049",
    journal = "Nucl. Phys. B Proc. Suppl.",
    volume = "207-208",
    pages = "188--191",
    year = "2010"
}

@article{MartinezTorres:2010ax,
    author = "Mart{\'i}nez Torres, A. and Garz{\'o}n, E. J. and Oset, E. and Dai, L. R.",
    title = "{Limits to the fixed center approximation to Faddeev equations: The case of the $\phi(2170)$}",
    doi = "10.1103/PhysRevD.83.116002",
    journal = "Phys. Rev. D",
    volume = "83",
    pages = "116002",
    year = "2011"
}

@article{Napsuciale:2007wp,
    author = "Napsuciale, M. and Oset, E. and Sasaki, K. and Vaquera-Araujo, C. A.",
    title = "{Electron-positron annihilation into $\phi f_{0}(980)$ and clues for a new $1^{--}$ resonance}",
    doi = "10.1103/PhysRevD.76.074012",
    journal = "Phys. Rev. D",
    volume = "76",
    pages = "074012",
    year = "2007"
}

@article{Zhao:2013ffn,
    author = "Zhao, Lu and Li, Ning and Zhu, Shi Lin and Zou, Bing Song",
    title = "{Meson-exchange model for the $\Lambda\bar{\Lambda}$ interaction}",
    doi = "10.1103/PhysRevD.87.054034",
    journal = "Phys. Rev. D",
    volume = "87",
    number = "5",
    pages = "054034",
    year = "2013"
}

@article{Deng:2013aca,
    author = "Deng, Chengrong and Ping, Jialun and Yang, Youchang and Wang, Fan",
    title = "{Baryonia and near-threshold enhancements}",
    doi = "10.1103/PhysRevD.88.074007",
    journal = "Phys. Rev. D",
    volume = "88",
    number = "7",
    pages = "074007",
    year = "2013"
}

@article{Dong:2017rmg,
    author = {Dong, Yubing and Faessler, Amand and Gutsche, Thomas and L{\"u}, Qi Fang and Lyubovitskij, Valery E.},
    title = "{Selected strong decays of $\eta(2225)$ and $\phi(2170)$ as $\Lambda \bar\Lambda$ bound states}",
    doi = "10.1103/PhysRevD.96.074027",
    journal = "Phys. Rev. D",
    volume = "96",
    number = "7",
    pages = "074027",
    year = "2017"
}

@article{Wei:2025ejv,
    author = "Wei, Xiang and Shen, Qing Hua and Liu, Xiao Hai and Xie, Ju Jun",
    title = "{Shedding light on the nature of the $\phi(2170)$ state in the $e^{+}e^{-}\to\phi\pi^{+}\pi^{-}$ reaction}",
    doi = "10.1103/frs3-zpt2",
    journal = "Phys. Rev. D",
    volume = "113",
    number = "1",
    pages = "014022",
    year = "2026"
}

@article{BaBar:2007ptr,
    author = "Aubert, Bernard and others",
    collaboration = "{BABAR Collaboration}",
    title = "{The $e^{+}e^{-}\to K^{+}K^{-}\pi^{+}\pi^{-}$, $K^{+}K^{-}\pi^{0}\pi^{0}$, and $K^{+}K^{-}K^{+}K^{-}$ cross sections measured with initial-state radiation}",
    reportNumber = "SLAC-PUB-12435, BABAR-PUB-07-021",
    doi = "10.1103/PhysRevD.76.012008",
    journal = "Phys. Rev. D",
    volume = "76",
    pages = "012008",
    year = "2007"
}

@article{BES:2007sqy,
    author = "Ablikim, Medina and others",
    collaboration = "{BES Collaboration}",
    title = "{Observation of $Y(2175)$ in $J/\psi\to\eta\phi f_0(980)$}",
    doi = "10.1103/PhysRevLett.100.102003",
    journal = "Phys. Rev. Lett.",
    volume = "100",
    pages = "102003",
    year = "2008"
}

@article{BESIII:2014ybv,
    author = "Ablikim, M. and others",
    collaboration = "{BESIII Collaboration}",
    title = "{Study of $J/\psi \to \eta \phi \pi^+ \pi^-$ at BESIII}",
    doi = "10.1103/PhysRevD.91.052017",
    journal = "Phys. Rev. D",
    volume = "91",
    number = "5",
    pages = "052017",
    year = "2015"
}

@article{BESIII:2017qkh,
    author = "Ablikim, Medina and others",
    collaboration = "{BESIII Collaboration}",
    title = "{Observation of $e^+ e^- \to \eta Y(2175)$ at center-of-mass energies above 3.7 GeV}",
    doi = "10.1103/PhysRevD.99.012014",
    journal = "Phys. Rev. D",
    volume = "99",
    number = "1",
    pages = "012014",
    year = "2019"
}

@article{GlueX:2025dgj,
    author = "Afzal, F. and others",
    collaboration = "{GlueX Collaboration}",
    title = "{Search for the $Y(2175)$ in the photoproduction cross-section measurement of $\gamma p\to\phi\pi^{+}\pi^{-}p$ at GlueX}",
    doi = "10.1103/jsfs-nq46",
    journal = "Phys. Rev. Lett.",
    volume = "136",
    number = "25",
    pages = "251902",
    year = "2026"
}

@article{ParticleDataGroup:2024cfk,
    author = "Navas, S. and others",
    collaboration = "Particle Data Group",
    title = "{Review of Particle Physics}",
    doi = "10.1103/PhysRevD.110.030001",
    journal = "Phys. Rev. D",
    volume = "110",
    number = "3",
    pages = "030001",
    year = "2024"
}

@misc{Bai:2026atm,
    author = "Bai, Zi Yue and Chen, Dian Yong and Huang, Qi and Liu, Xiang and Luo, Si Qiang and Wang, Jun Zhang",
    title = "{Unquenched Charmonium and Beyond}",
    eprint = "2602.19887",
    archivePrefix = "arXiv",
    primaryClass = "hep-ph",
    year = "2026"
}

@article{Wang:2016qmz,
    author = "Wang, Bo and Liu, Xiang and Chen, Dian Yong",
    title = "{Prediction of anomalous $\Upsilon(5S)\to\Upsilon(1^3D_J)\eta$ transitions}",
    doi = "10.1103/PhysRevD.94.094039",
    journal = "Phys. Rev. D",
    volume = "94",
    number = "9",
    pages = "094039",
    year = "2016"
}

@article{Belle:2018hjt,
    author = "Tamponi, U. and others",
    collaboration = "{Belle Collaboration}",
    title = "{Inclusive study of bottomonium production in association with an $\eta$ meson in $e^{+}e^{-}$ annihilations near $\Upsilon(5S)$}",
    reportNumber = "BELLE-PREPRINT-2018-01, KEK-PREPRINT-2017-60",
    doi = "10.1140/epjc/s10052-018-6086-4",
    journal = "Eur. Phys. J. C",
    volume = "78",
    number = "8",
    pages = "633",
    year = "2018"
}

@article{Kaymakcalan:1983qq,
    author = "Kaymakcalan, {\"O}. and Rajeev, S. and Schechter, J.",
    title = "{Non-Abelian anomaly and vector-meson decays}",
    reportNumber = "SU-4222-278, COO-3533-278",
    doi = "10.1103/PhysRevD.30.594",
    journal = "Phys. Rev. D",
    volume = "30",
    pages = "594",
    year = "1984"
}

@article{Zhou:2022wwk,
    author = "Zhou, Qin Song and Wang, Jun Zhang and Liu, Xiang",
    title = "{Role of the $\omega(4S)$ and $\omega(3D)$ states in mediating the $e^{+}e^{-}\to\omega\eta$ and $\omega\pi^{0}\pi^{0}$ processes}",
    doi = "10.1103/PhysRevD.106.034010",
    journal = "Phys. Rev. D",
    volume = "106",
    number = "3",
    pages = "034010",
    year = "2022"
}

@article{Zhou:2022ark,
    author = "Zhou, Qin Song and Wang, Jun Zhang and Liu, Xiang and Matsuki, Takayuki",
    title = "{Identifying the contribution of higher $\rho$ mesons around 2~GeV in the $e^{+}e^{-}\to\omega\pi^{0}$ and $e^{+}e^{-}\to\rho\eta^{\prime}$ processes}",
    doi = "10.1103/PhysRevD.105.074035",
    journal = "Phys. Rev. D",
    volume = "105",
    number = "7",
    pages = "074035",
    year = "2022"
}

@article{Locher:1993cc,
    author = "Locher, M. P. and Lu, Y. and Zou, B. S.",
    title = "{Rates for the reactions $\bar{p}p\to\pi\phi$ and $\gamma\phi$}",
    reportNumber = "PSI-PR-93-20",
    doi = "10.1007/BF01289796",
    journal = "Z. Phys. A",
    volume = "347",
    pages = "281--284",
    year = "1994"
}

@article{Li:1996yn,
    author = "Li, Xue Qian and Bugg, David V. and Zou, Bing Song",
    title = "{Possible explanation of the ``$\rho\pi$ puzzle'' in $J/\psi$, $\psi^{\prime}$ decays}",
    reportNumber = "RAL-TR-96-062",
    doi = "10.1103/PhysRevD.55.1421",
    journal = "Phys. Rev. D",
    volume = "55",
    pages = "1421--1424",
    year = "1997"
}

@article{Cheng:2004ru,
    author = "Cheng, Hai Yang and Chua, Chun Khiang and Soni, Amarjit",
    title = "{Final-state interactions in hadronic $B$ decays}",
    reportNumber = "BNL-HET-04-17",
    doi = "10.1103/PhysRevD.71.014030",
    journal = "Phys. Rev. D",
    volume = "71",
    pages = "014030",
    year = "2005"
}

@article{Wang:2022jxj,
    author = "Wang, Jun Zhang and Liu, Xiang",
    title = "{Confirming the existence of a new higher charmonium $\psi(4500)$ by the newly released data of $e^{+}e^{-}\to K^{+}K^{-}J/\psi$}",
    doi = "10.1103/PhysRevD.107.054016",
    journal = "Phys. Rev. D",
    volume = "107",
    number = "5",
    pages = "054016",
    year = "2023"
}

@article{Peng:2024xui,
    author = "Peng, Tian Cai and Bai, Zi Yue and Wang, Jun Zhang and Liu, Xiang",
    title = "{How higher charmonia shape the puzzling data of the $e^{+}e^{-}\to\eta J/\psi$ cross section}",
    doi = "10.1103/PhysRevD.109.094048",
    journal = "Phys. Rev. D",
    volume = "109",
    number = "9",
    pages = "094048",
    year = "2024"
}

@article{Qian:2023taw,
    author = "Qian, Ri Qing and Liu, Xiang",
    title = "{Production of charmonium $\chi_{cJ}(2P)$ plus one $\omega$ meson by $e^+e^-$ annihilation}",
    doi = "10.1103/PhysRevD.108.094046",
    journal = "Phys. Rev. D",
    volume = "108",
    number = "9",
    pages = "094046",
    year = "2023"
}

@article{Gilman:1987ax,
    author = "Gilman, Frederick J. and Kauffman, Russel",
    title = "{$\eta$-$\eta^{\prime}$ mixing angle}",
    reportNumber = "SLAC-PUB-4301",
    doi = "10.1103/PhysRevD.36.2761",
    journal = "Phys. Rev. D",
    volume = "36",
    pages = "2761",
    year = "1987",
    note = "[Erratum: Phys. Rev. D 37, 3348 (1988)]"
}

@article{Christ:2010dd,
    author = "Christ, N. H. and Dawson, C. and Izubuchi, T. and Jung, C. and Liu, Q. and Mawhinney, R. D. and Sachrajda, C. T. and Soni, A. and Zhou, R.",
    title = "{$\eta$ and $\eta^{\prime}$ Mesons from Lattice QCD}",
    doi = "10.1103/PhysRevLett.105.241601",
    journal = "Phys. Rev. Lett.",
    volume = "105",
    pages = "241601",
    year = "2010"
}

@article{Yu:2025pyu,
    author = "Yu, Zhuo and Wu, Qi and Chen, Dian Yong",
    title = "{$Z_{cs}^{+}$ production in $B^{+}$ decays}",
    doi = "10.1103/fjsv-fxpm",
    journal = "Phys. Rev. D",
    volume = "112",
    number = "9",
    pages = "094054",
    year = "2025"
}

\end{document}